\documentclass[a4paper,11pt]{article}

\usepackage{mathtools,bbm}  
\usepackage{mathrsfs}
\usepackage{amsmath}
\usepackage{amssymb}
\usepackage{textcomp}
\usepackage{url}
\usepackage{color}
\usepackage{cite}
\usepackage{array}  
\usepackage[hyperfootnotes=false, linktocpage=true, colorlinks, citecolor=blue, linkcolor=blue, urlcolor=blue]{hyperref}

\numberwithin{equation}{section}

\newlength{\xtrawidth}
\newlength{\xtraheight}
\newcommand{\kahl}{K\"ahler}
\newcommand{\be}{\begin{equation}}
\newcommand{\ee}{\end{equation}}
\newcommand{\bea}{\begin{eqnarray}}
\newcommand{\eea}{\end{eqnarray}}
\newcommand{\bi}{\begin{itemize}}
\newcommand{\ei}{\end{itemize}}

\newcommand{\mc}{\mathcal}
\newcommand{\mbb}{\mathbb}

\newcommand{\h}{\hat}

\newcommand{\cohomCalg}{{\fontfamily{put}\footnotesize\selectfont cohomCalg}~}

\renewcommand{\P}{\mathbb{P}} 
\newcommand{\C}{\mathbb{C}}

\newcolumntype{C}{>{$}c<{$}}

\newcolumntype{W}[1]{>{\centering\arraybackslash$}m{#1}<{$}} 
\newlength{\mycolwd}
\pdfoutput=1

\begin{document}
\begin{centering}
\vspace*{1.2cm}
{\Large \bf  Heterotic Line Bundle Models on Elliptically Fibered\\[2mm] Calabi-Yau Three-folds}

\vspace{1cm}

{\bf Andreas P.~Braun}\footnote{andreas.braun@physics.ox.ac.uk},
{\bf Callum R. Brodie}\footnote{callum.brodie@physics.ox.ac.uk},
{\bf Andre Lukas}\footnote{lukas@physics.ox.ac.uk}\\
{\small
\vspace*{.5cm}
Rudolf Peierls Centre for Theoretical Physics, University of Oxford\\
  1 Keble Road, Oxford OX1 3NP, UK\\[1cm]
}
\begin{abstract}\noindent
We analyze heterotic line bundle models on elliptically fibered Calabi-Yau three-folds over weak Fano bases. In order to facilitate Wilson line breaking to the standard model group, we focus on elliptically fibered three-folds with a second section and a freely-acting involution. Specifically, we consider toric weak Fano surfaces as base manifolds and identify six such manifolds with the required properties. The requisite mathematical tools for the construction of line bundle models on these spaces, including the calculation of line bundle cohomology, are developed. A computer scan leads to more than $400$ line bundle models with the right number of families and an $SU(5)$ GUT group which can descend to standard-like models after taking the $\mathbb{Z}_2$ quotient. A common and surprising feature of these models is the presence of a large number of vector-like states.
\end{abstract}
\end{centering}

\newpage

\tableofcontents
\setcounter{footnote}{0}
\newpage

\section{Introduction}
Heterotic string model building on Calabi-Yau (CY) manifolds requires an understanding of the gauge bundle on the compactification space and this constitutes one of the main technical challenges in constructing realistic models, particularly for gauge bundles with a non-Abelian structure group. Indeed, only a relatively small number of quasi-realistic models based on such gauge bundles are known in the literature~\cite{Braun:2005ux,Bouchard:2005ag,Anderson:2009mh,Braun:2011ni}.

However, it has been realised~\cite{Anderson:2011ns,Anderson:2012yf} that quasi-realistic models can also be constructed using gauge bundles with Abelian structure groups which are considerably easier to handle than their non-Abelian counterparts. For these models, the vector bundle is a direct sum of line bundles which are classified and can be analyzed systematically.  By scanning this space of heterotic line bundle models, a large number of examples which lead to the standard model spectrum has been found~\cite{Anderson:2011ns,Anderson:2012yf}. For technical reasons, these models have been obtained using the arguably simplest type of CY manifolds - complete intersections in product of projective spaces - and, to a lesser extent, CY hypersurfaces in toric four-folds. 
The main purpose of the present paper is to study heterotic line bundle models for another class of CY manifolds, namely elliptically fibered CY three-folds. 

There are a number of motivations for addressing this problem. Firstly, we would like to develop the necessary tools for constructing line bundle models on elliptically fibered CY three-folds, including the calculation of line bundle cohomology on these manifolds. Secondly, by studying line bundle models on another class of CY manifolds, we would like to gain some insight into which of their properties are generic and which are related to the particular type of underlying CY manifold. Finally, we are motivated by heterotic F-theory duality which is formulated for elliptically fibered CY manifolds. More specifically, this duality normally relies on spectral cover bundles~\cite{Friedman:1997yq}, usually with a non-Abelian structure group, on elliptically fibered CY manifolds. It would be interesting to understand in detail how the duality works for line bundle models. In the present paper we focus on the construction of models on the heterotic side as a first step in this direction while the discussion of heterotic F-theory duality for these models will be the subject of future work. 

There are strong indications~\cite{Anderson:2014hia} that the construction of phenomenologically interesting heterotic CY models requires a CY manifold with a non-trivial first fundamental group which facilitates the presence of a Wilson line. In such constructions, the part of the bundle with a non-flat connection is used to break to a GUT group, such as $SU(5)$, while the Wilson line breaks further to the standard model group. The correct chiral asymmetry is easily engineered at the GUT level by imposing a single condition on the index of the bundle. The Wilson line, while breaking the GUT multiplets up into standard model multiplets, does not change this chiral asymmetry. In this way three chiral families are easily obtained by imposing one condition on the bundle. On the other hand, the direct breaking to the standard model group without the presence of a Wilson line requires imposing one index condition per standard model multiplet. It can be shown~\cite{Anderson:2014hia} that the combination of these conditions leads to serious phenomenological problems. 

For this reason we would like to follow the standard two-step construction with an intermediate GUT group, which we choose to be $SU(5)$, and subsequent Wilson line breaking. The required CY manifolds with a non-trivial first fundamental group are usually constructed by starting with a simply-connected ``upstairs" CY manifold $X$ with a freely-acting discrete symmetry $\Gamma$ and then taking the quotient $X/\Gamma$. We would like to follow this approach and should, therefore, construct elliptically fibered CY three-folds with freely-acting symmetries. For our purpose, we consider the simplest case where $\Gamma=\mathbb{Z}_2$, that is, we consider freely-acting involutions. The construction and analysis of such elliptically fibered CY three-folds with a freely-acting involution constitutes another technical complication, partly addressed in earlier work~\cite{Donagi:1999ez,Andreas:1999ty}, which we review and further develop for our purposes. 

The main idea for constructing such freely-acting involutions is as follows. Consider an elliptically fibered CY three-fold $X$ with projection $\pi:X\rightarrow B$, two-fold base $B$  and with a section $\sigma:B\rightarrow X$. We can describe the typical elliptic fiber, $E_b=\pi^{-1}(b)$ where $b\in B$, by its Jacobian, that is, by a complex $w$ plane with identifications $w\sim w+1$ and $w\sim w+\tau_b$, where $\tau_b$ is the complex structure of $E_b$. An obvious starting point is to consider an involution $\iota_X$ of $X$ which acts on the fiber as a half-shift, that is, as $\iota_E:w\rightarrow w+1/2$. This looks promising since this action of $\iota_E$ is already fixed point free. However, there are two complications which can be inferred from the following argument. The presence of the half-shift on the fibers means that the elliptic fibration has to have a second section, $\zeta=\iota_E\circ\sigma$, in addition to $\sigma$. Such elliptic fibrations with two sections are known to have a special structure and, in particular, the discriminant locus (the locus on the base $B$ over which the torus fiber degenerates) splits into two components. It turns out that on one of these components the fiber degenerates such that the entire CY manifold becomes singular. To produce a smooth manifold, the singular loci need to be blown-up and this leads to an additional (second and fourth) cohomology class. The other component of the discriminant locus does not lead to singularities of $X$  but  the action of $\iota_E$ on the corresponding degenerate fibers is not fixed point free. To remedy this problem we construct $\iota_X$ by  combining the half-shift, $\iota_E$, on the fiber with a simultaneous action $\iota_B$ on the base. The latter does not have to be fixed point free on $B$. In order to ensure that $\iota_X$ is fixed point free it is sufficient to require that the fixed point locus of $\iota_B$ should not intersect the above mentioned second component of the discriminant locus where $\iota_E$ has fixed points on the fiber. This is generically the case if $\iota_B$ has at most fixed points (rather than fixed curves) on $B$ and this is what we will require. 

This construction has implications for the possible choices of  base spaces $B$. Smoothness of a generic Weierstrass model over $B$ implies that $B$ should be weak Fano~\cite{Morrison:2012js,Halverson:2015jua}. Further, finding a freely-acting involution $\iota_X$ along the lines described above requires the existence of an involution $\iota_B$ with at most fixed points. We will focus on two-dimensional toric Fano base spaces $B$ and find that there are six such spaces which lead to smooth CY three-folds and have a suitable involution $\iota_B$.

For such elliptically fibered three-folds $X$ with a freely-acting involution and two sections we systematically develop the required model building tools. This includes the construction of a suitable integral basis of the second and fourth homology, calculation of the intersection ring, the construction of K\"ahler and Mori cones and the analysis of line bundles $L\rightarrow X$ and their properties. In particular, we show how to calculate the cohomology of line bundles by combining the Leray spectral sequence with known methods for calculating line bundle cohomology on toric manifolds~\cite{CohomOfLineBundles:Algorithm,cohomCalg:Implementation}.

Based on these results, we scan rank five line bundle models for the six suitable base spaces and find more than $400$ $SU(5)$ GUT models with the correct chiral asymmetry. Upon taking the quotient by the involution and introducing a $\mathbb{Z}_2$ Wilson line in the hypercharge direction these will lead to models with the standard model gauge group and three chiral families. A common feature of all models is the presence of a large number of vector-like states. In $SU(5)$ language all models have at least one ${\bf 10}$--$\overline{\bf 10}$ vector-like pair and at least $20$ (!) ${\bf 5}$--$\bar{\bf 5}$ vector-like pairs. Particularly the latter number is surprising compared with the results obtained for heterotic line bundle models on complete intersection CY manifolds~\cite{Anderson:2011ns,Anderson:2012yf,Anderson:2013xka}. For complete intersection CY manifolds, imposing the right chiral asymmetry frequently meant the absence of ${\bf 10}$--$\overline{\bf 10}$ vector-like pairs and a small number of ${\bf 5}$--$\bar{\bf 5}$ vector-like pairs. 

The plan of the paper is as follows. In the next section, we will introduce elliptically fibered CY manifolds and discuss their properties including line bundles on these spaces. In Section~\ref{line_bundle_models}, heterotic line bundle models are reviewed and in Section~\ref{sec_scan} we present the results of our model scan. We conclude in Section~\ref{sec_con}. In the main part of the paper, we will keep the discussion informal and as non-technical as the subject allows; technical details can be found in the Appendices.

\section{Elliptically fibered CY three-folds and involutions}
In this section, we discuss the required background on elliptically fibered CY three-folds, specifically in the presence of a freely-acting involution. Much of the material is not new and can be inferred by combining results in the literature~\cite{Donagi:1999ez,Andreas:1999ty} but we would like to present a coherent exposition as required for systematic model building. This will be kept somewhat informal with focus on the main results  while technical details can be found in
Appendices~\ref{ell_curve} and \ref{bases_and_involutions}.

\subsection{Elliptically fibered CY three-folds with a single section}
\label{fibrsingsec_maintext}
As a warm-up, we consider elliptically fibered Calabi-Yau three-folds $X$ over a two-fold base $B$ with projection $\pi:X\rightarrow B$ and a single section $\sigma:B\rightarrow X$. Points on the base are denoted by $b\in B$ and $K_B$ is the canonical bundle of the base. A Weierstrass model for $X$ is given by the equation
\be\label{eq:generic_weierstrass}
 zy^2=x^3+f(b)xz^2+g(b)z^3\, ,
\ee
which, for each base point $b\in B$, describes an elliptic curve, $E_b=\pi^{-1}(b)$, embedded in $\mbb{P}^2$ with homogeneous coordinates $x$, $y$, $z$. Here, $f$ and $g$ are sections of the line bundles $K_B^{-4}$ and $K_B^{-6}$, respectively, which encode the variation of the elliptic curve over the base. 
In these $\mbb{P}^2$ coordinates the section can be explicitly written as $\sigma(b)=(b,(0,1,0))$, that is, it is ``located'' at the point $(x,y,z)=(0,1,0)\in E_b$ in each fiber. 

A typical torus fiber $E_b$ can also be described by its Jacobian, that is, by a complex $w$ plane with identifications $w\sim w+1$ and $w\sim w+\tau_b$, where $\tau_b$ is the complex structure of $E_b$. For each point $b\in B$, this complex structure is related to the sections $f$ and $g$ by the equation
\be
j(\tau_b) = \frac{4(24f(b))^3}{4f(b)^3+27g(b)^2} \; .
\ee
The denominator on the right-hand side of this expression, that is,
\begin{equation}
 \Delta=4f^3+27 g^2 \label{Deltadef}
\end{equation}
is a section of $K_B^{-12}$ and is called the discriminant. The discriminant locus, defined by $\{b\in B\;|\;\Delta(b)=0\}$, is a curve in the base over which the fiber $E_b$ degenerates. This will be discussed in more detail in the next sub-section for the case of two sections.

A basis of curves and divisors of $X$ can be obtained from a basis of curves on $B$ by using the maps $\pi^{-1}$ and $\sigma$. Other relevant properties of $X$, such as the intersection ring, the Mori cone and the K\"ahler cone, can also be obtained in terms of properties of the base. Since we are primarily interested in the two-section case, we will defer the details of this to the following sub-sections.

\subsection{Elliptically fibered CY three-folds with two sections}
\label{fibrtwosec_maintext}
As indicated earlier, we would like to construct elliptically fibered CY three-folds $X$ with a freely-acting involution $\iota_X$, starting with an involution $\iota_E$ on each fiber which acts as a half-shift $w\rightarrow w+1/2$. The presence of $\iota_E$ implies the presence of a second section $\zeta=\iota_E\circ\sigma:B\rightarrow X$ of the fibration, located at the ``two-torsion point" in each fiber, in addition to the section $\sigma:B\rightarrow X$. We should, therefore, discuss the structure of elliptically fibered CY three-folds with two sections.

Starting from the general Weierstrass model~\eqref{eq:generic_weierstrass}, a Weierstrass model with a second section can be found by an appropriate tuning of parameters, the general form of which has been found in Ref.~\cite{Morrison:2012ei}. As explained in Appendix \ref{fibrtwosec_appendix}, a second section located at the two-torsion point of the elliptic fiber emerges by choosing the specific, factorisable form (see also Ref.~\cite{Donagi:1999ez})
\be\label{eq:twosectionweier}
zy^2=(x-\alpha z)(x^2+\alpha xz+\beta z^2)\, ,
\ee
of the Weierstrass model, where $\alpha$ and $\beta$ are sections of $K_B^{-2}$ and $K_B^{-4}$, respectively. Comparison with Eq.~\eqref{eq:generic_weierstrass} shows that this corresponds to choosing the sections $f$ and $g$ in the general Weierstrass model as
\be
f = \beta - \alpha^2 \,, \hspace{1cm} g = -\alpha\beta \, .
\ee
The two sections of such a fibration are given by
\be
\sigma(b)=(b,(0,1,0))\,, \quad \zeta(b)=(b,(\alpha,0,1))\; ,
\ee
so they are located at $(x,y,z)=(0,1,0)$ and $(x,y,z)=(\alpha,0,1)$, respectively.

As can be inferred from the factorised form of the Weierstrass model, this model has singularities which we need to resolve. This can be explicitly seen be working out the discriminant~\eqref{Deltadef} which becomes
\be
\Delta =  \Delta_1 \Delta_2^2\,,\quad  \textrm{where} \quad \Delta_1 := 4\beta-\alpha^2\,, \quad \Delta_2 := 2\alpha^2+\beta\; . \label{Delta2}
\ee
Using this notation and (shifted) affine coordinates $X=x/z-\alpha$, $Y=y/z$ on the patch of $\mathbb{P}^2$ where $z\neq 0$, it is shown in Appendix~\ref{fibrtwosec_appendix} that the half-shift $\iota_E$ on regular fibers can be written as 
\be
X\rightarrow  \frac{\Delta_2}{X} \,, \quad Y\rightarrow-\frac{\Delta_2Y}{X^2} \;.  \label{halfshift}
\ee
From the above discriminant, there are singularities, corresponding to fibers of Kodaira type $I_2$, over the curve $\{b\in B\,|\,\Delta_2(b)=0\}$ in the base. In order to obtain a smooth CY three-fold, these singularities need to be resolved, after which the fibers over the locus $\Delta_2(b)=0$ become a pair of $\mathbb{P}^1$'s touching in a point. In addition, the action of $\iota_E$ has to be extended to these blown-up fibers. Appendix~\ref{fibrtwosec_appendix} provides the details of the blow-up procedure and shows explicitly that $\iota_E$ can indeed be extended to the blow-up and has no fixed points on the blown-up fibers. More specifically, it turns out that $\iota_E$ exchanges the two $\mathbb{P}^1$ curves of the blown-up fibers. For these reducible fibers, the section $\sigma$ takes values in one of the $\mathbb{P}^1$ curves and the second section $\zeta$ takes values in the other $\mathbb{P}^1$.

From Eq.~\eqref{Delta2}, there is another component of the discriminant locus, given by the curve $\{b\in B\,|\,\Delta_1(b)=0\}$ in the base. The degenerate fibers over this locus do not lead to singularities of the entire three-fold so that there is no need for a further blow-up. However, as discussed in Appendix~\ref{fibrtwosec_appendix} , the map $\iota_E$ does have fixed points on the fibers over this locus, while it is fixed point free on all other fibers.  This means that the involution $\iota_X$, if it is to be fixed point free, cannot simply be taken to be $\iota_E$ with a trivial action on the base. However, we can obtain a fixed point free involution $\iota_X$ by combining $\iota_E$ with an involution $\iota_B$ on the base whose fixed point locus does not intersect the curve $\{b\in B\,|\,\Delta_1(b)=0\}$. This is generically the case if $\iota_B$ has at most fixed points (rather than fixed curves) on $B$ and this is what we will require. The presence of such an involution $\iota_B$ places constraints on the allowed base manifolds which we will discuss below. 

Alternatively, one may realise elliptic Calabi-Yau three-folds with a second section at the two-torsion point by embedding the elliptic fiber into the Hirzebruch surface $\mathbb{F}_2$, see \cite{Grimm:2010ez,Morrison:2012ei,Mayrhofer:2012zy}. We will review this approach in Appendix \ref{app:toric2sectionmodels}. This approach has the advantage that, for a toric base space $B$, the corresponding elliptic Calabi-Yau three-fold can be presented as a hypersurface in a toric variety. We will use this alternative method to realise the relevant CY manifolds to check some of our results, particularly in relation to line bundle cohomology.

\subsection{Curves and divisors}
We would now like to construct a number of objects, as required for our model building purposes, including a basis of curve and divisor classes, on elliptically fibered three-folds $X$ with a freely-acting involution and two sections, in terms of the corresponding objects on the base $B$. Technical details can be found in Appendix~\ref{app:intersectionson3fold}.

We begin by introducing an integral basis $\{{\cal C}^i\}$ of the second homology of the base $B$, as well as a dual basis, $\{{\cal C}_i\}$, of curve classes such that
\begin{equation}
 \mc{C}^i\cdot \mc{C}_j=\delta^i_j\; .
\end{equation} 
Here and in the following we use indices $i,j,\ldots$ to indicate the index range $1,\ldots ,h^{1,1}(B)$. The intersection forms for these basis curves are denoted by
\begin{equation}
 g_{ij}:=\mc{C}_i\cdot\mc{C}_j\;,\qquad g^{ij}:=\mc{C}^i\cdot\mc{C}^j\; ,
\end{equation}
and it is easy to see that $g^{ij}$ is the inverse of $g_{ij}$. Further, these ``metrics" can be used to raise and lower indices, that is, ${\cal C}^i=g^{ij}{\cal C}_j$ and ${\cal C}_i=g_{ij}{\cal C}^j$. For later purposes, it is also useful to introduce the quantities
\begin{equation}
 \lambda_i := K_B\cdot\mc{C}_i \,, \quad \lambda := K_B^2=c_1(B)^2 = \lambda^i\lambda_i\; ,
\end{equation}
where $K_B=-c_1(B)$ is the canonical bundle of the base and $\lambda^i=g^{ij}\lambda_j$.

We can use either one of the two sections $\sigma$ and $\zeta$ to raise these curve classes on the base to curve classes on the entire three-fold. The three-fold has two more classes which cannot be obtained in this way. These are the class $F$ of a generic fiber and the new class $N$ introduced by the blow-up, the latter chosen such that the component of the reducible fibers over $\{b\in B\,|\,\Delta_2(b)=0\}$ which is met by the zero section $\sigma$ has class $F-N$. For a general curve class, ${\cal C}$,  on the base we have the relation
\begin{equation}
 \zeta({\cal C})=\sigma({\cal C})+({\cal C}\cdot c_1(B))\left[F-2N\right]\; ,
\end{equation} 
which shows that lifts of base curves with the two sections $\sigma$ and $\zeta$ are linearly related. It is, therefore, sufficient to consider lifts by one of the sections and we will use the zero section $\sigma$ for this purpose. Accordingly, we introduce a basis $\{C^I\}$, where $I=(0,\hat{0},i)$, of curve classes on $X$ by
\begin{equation}
 C^{0}=F-N\;,\quad C^{\hat{0}}=N\;,\quad C^i=\sigma({\cal C}^i)-\lambda^i(F-N)\; . \label{Cbasis}
\end{equation} 

Divisor classes on $X$ can be obtained from curve classes, ${\cal C}$, on the base by the inverse image $\pi^{-1}({\cal C})$. There are two further classes, the images $\sigma(B)$ and $\zeta(B)$ of the base under the two sections, which cannot be obtained in this way. Hence, we introduce a basis $\{D_I\}$ of divisor classes, where $I=(0,\hat{0},i)$, by
\begin{equation}
D_{0}=\sigma(B)\;,\quad D_{\hat{0}}=\zeta(B)\;,\quad D_i=\pi^{-1}({\cal C}_i)\; . \label{Dbasis}
\end{equation} 
From the intersections in Table~\ref{tab_CDisec} in Appendix~\ref{app:intersectionson3fold} we can see that this basis of divisor classes is dual to the above basis of curve classes, that is
\begin{equation}
 D_I\cdot C^J=\delta_I^J\; .
\end{equation}
Combining the information from Table~\ref{tab_CDisec} and Table~\ref{tab_DDisec} in Appendix~\ref{app:intersectionson3fold} we can work out the intersection numbers
\begin{equation}
d_{IJK}=D_I\cdot D_J\cdot D_K
\end{equation}
which are explicitly given by
\begin{equation}
d_{000}=d_{\hat{0}\hat{0}\hat{0}}=\lambda\;, \quad d_{00i}=d_{\hat{0}\hat{0}i}=\lambda_i \;, \quad d_{0ij}=d_{\hat{0}ij}=g_{ij}\;, \label{isec1}
\end{equation}
with all other components either fixed by symmetry from the above or else vanishing. 

For our model building purposes we also require the Mori and K\"ahler cones of $X$. We begin with the Mori cone, ${\cal M}_X$ of $X$ which, following Appendix~\ref{mori_kahl_cones_appendix}, can be written as
\begin{equation}
 {\cal M}_X=\left\{n_0(F-N)+n_{\hat{0}}N+\sigma({\cal C})+\zeta({\cal C}')\,|\,n_0,n_{\hat{0}}\in\mathbb{Z}^{\geq 0}\;,\;\; {\cal C},{\cal C}'\in{\cal M}_B\right\}\; , \label{MoriX}
\end{equation}
where ${\cal M}_B$ is the Mori cone of the base. If we write K\"ahler forms as $J=t^ID_I$ with the K\"ahler moduli ${\bf t}=(t^I)$ relative to the divisor basis $\{D_I\}$, the K\"ahler cone is the dual of the Mori cone, that is,
\begin{eqnarray} \label{Kcone}
 {\cal K}_X&=&\left\{J=t^ID_I\,|\,J\cdot C\geq 0\mbox{ for all } C\in{\cal M}_X\right\}\\
 &\cong&\left\{{\bf t}\,|\, {\bf t}\cdot{\bf n}\geq 0\mbox{ for all }n^IC_I\in{\cal M}_X\right\}\; .
 \end{eqnarray} 
and can, hence, be explicitly worked out once the Mori cone is known.

We should also discuss how the above basis of curve and divisor classes relates to the involution $\iota_X$ on $X$ and its action $\iota_B$ on the base. We find
\begin{align}
 &\iota_X(F-N)=N\;,\quad \iota_X(\sigma({\cal C}))=\zeta(\iota_B({\cal C}))\; ,\\
 &\iota_X(\sigma(B))=\zeta(B)\;,\quad \iota_X(\pi^{-1}({\cal C}))=\pi^{-1}(\iota_B({\cal C}))\; ,\label{invD}
\end{align} 
where ${\cal C}$ is a curve class on the base. In particular, as is evident from the second line, $\iota_X$ exchanges $D_0=\sigma(B)$ with $D_{\hat{0}}=\zeta(B)$.

Finally, the second Chern class and Euler number of $X$ are given by (see Refs.~\cite{Andreas:1999ty,Donagi:1999ez})
\begin{eqnarray}
c_2(X) &=&12\sigma(c_1(B))+(c_2(B)+11c_1(B)^2)(F-N)+(c_2(B)-c_1(B)^2)N\label{c2X}\\
&=& (c_2(B)-\lambda)(C^0+C^{\hat{0}})-12\lambda_iC^i\label{c2XC} \; , \\
\chi(X)&=&-36\int_Bc_1(B)^2=-36\lambda\; . \label{chiX}
\end{eqnarray}
We have thus succeeded in expressing all basic properties of elliptically fibered CY three-folds with two sections in terms of corresponding properties of the base. 
 
\subsection{Line bundles} \label{sec:lbs}
We would now like to collect properties of line bundles $L\rightarrow X$ on elliptically fibered CY three-folds with a freely-acting involution and two sections. As usual, we denote by ${\cal O}_X(D)$ a line bundle with first Chern class or character
\begin{equation}
 {\rm ch}_1({\cal O}_X(D))=c_1({\cal O}_X(D))=D\; .
\end{equation} 
To write down explicit expressions for the second Chern character and the index of a line bundle it is convenient to represent the corresponding divisor as a linear combination $D=k^ID_I$, where $k^I\in\mathbb{Z}$, relative to the basis $\{D_I\}$. Then we find
\begin{eqnarray}
 {\rm ch}_2({\cal O}_X(k^ID_I))&=&\frac{1}{2}d_{IJK}k^Ik^JC^K\\
 {\rm ind}({\cal O}_X(k^ID_I))&=&\frac{1}{6}d_{IJK}k^Ik^Jk^K+\frac{1}{12}k^Ic_{2I}(X)\; , \label{indL}
\end{eqnarray} 
where $c_{2I}(X)$ are the components of the second Chern class of $X$, defined by $c_2(X)=c_{2I}(X)C^I$, given in Eq.~\eqref{c2XC}. We also require an expression for the slope of a line bundle. With the K\"ahler form written as $J=t^ID_I$ the slope of a line bundle $L={\cal O}_X(k^ID_I)$ is defined by
\begin{equation}
\mu_X(L):=\int_XJ\wedge J\wedge c_1(L)=d_{IJK}t^It^Jk^K\; .
\end{equation} 
Later, we will be interested in line bundles $L$ whose slope vanishes somewhere in the (interior, $\mathring{\cal K}_X$, of the) K\"ahler cone~\eqref{Kcone}, so we have to solve the quadratic equation $d_{IJK}t^It^Jk^K=0$ for ${\bf t}\in\mathring{\cal K}_X$.

To build heterotic line bundle models, we also need to know which line bundles $L$ admit an equivariant structure under the involution $\iota_X$. In fact, in this paper, we will be content checking invariance of $L$, a necessary but not sufficient condition for equivariance which is better suited for a systematic model search. A line bundle $L={\cal O}_X(D)$ is invariant under $\iota_X$ iff $\iota_X^*L\cong L$ or, equivalently, iff $\iota_X(D)=D$. From Eqs.~\eqref{invD}, the last condition can be worked out more explicitly as
\begin{equation}
 \iota_X(k^ID_I)=k^ID_I\quad\Longleftrightarrow\quad k^0=k^{\hat{0}}\mbox{ and } I^i_{Bj}k^j=k^i\; , \label{Dinv}
\end{equation} 
where $I_B$ is a matrix which describes the action of the involution $\iota_B$ on the basis $\{{\cal C}_i\}$ of curve classes on the base such that
\begin{equation}
 \iota_B({\cal C}_j)=I^i_{Bj}{\cal C}_i\; . \label{Idef}
\end{equation}
Stated differently, invariant line bundles must be of the form
\begin{equation}
 L={\cal O}_X(n\Sigma)\otimes\pi^*({\cal L})\; ,\quad \Sigma=\sigma(B)+\zeta(B)\; , \label{lbinv}
\end{equation}
where $n\in\mathbb{Z}$ and ${\cal L}={\cal O}_B(k^i{\cal C}_i)$ is a line bundle on the base which satisfies $I^i_{Bj}k^j=k^i$.

To determine the full spectrum of line bundle models we need to compute line bundle cohomology, rather than merely line bundle indices. Later on, we will be interested in line bundles $L\rightarrow X$ whose slope $\mu_X(L)$ vanishes somewhere in the interior of K\"ahler cone. It is useful to note that, from a general vanishing theorem~\cite{kobayashi1986differential}, such line bundles satisfy
\begin{equation}
 h^0(X,L)=h^3(X,L)=0\quad\Longrightarrow\quad {\rm ind}(L)=h^2(X,L)-h^1(X,L)\; ,
\end{equation} 
with the exception of the trivial line bundle. Even for such line bundles, the index does not provide the full information and at least one more cohomology needs to be computed. 

The cohomology of line bundles $L\rightarrow X$ on an elliptically fibered CY three-fold $X$ can be expressed in terms of line bundle cohomology on the base $B$, using the direct image $\pi_*L$ and the associated higher direct images, $R^q\pi_*L$, together with the Leray spectral sequence~\cite{Donagi:2004ia,Friedman:1997yq,Andreas:2007ev}. Here we outline the structure of this calculation and its main results. Further details can be found in Appendix \ref{linebundles_cohomologies_appendix}, including details on the Leray spectral sequence and computations of the direct images and higher direct images.

The starting point of the computation is the exact sequence
\be
0 \to E_2^{1,0} \to H^1 \to E_2^{0,1} \to E_2^{2,0} \to H^2 \to E_2^{1,1} \to 0 \,, \label{leray1}
\ee
where
\be
H^p := H^p(X,L)\,, \quad E_2^{p,q}:= H^p(B,R^q\pi_*L)\; ,
\ee
which follows from the Leray spectral sequence. For our model building effort we require invariant line bundles, that is line bundles of the form~\eqref{lbinv}, and we will restrict the cohomology calculation to such cases. To work out the higher direct images of such invariant line bundles the formula $R^i\pi_*({\cal O}_X(n\Sigma)\otimes\pi^*({\cal L}))=R^i\pi_*({\cal O}_X(n\Sigma))\otimes{\cal L}$ is of some help and shows that all we require is the (higher) direct images of the line bundles ${\cal O}_X(n\Sigma)$. These are explicitly given by
\be
\pi_*\mc{O}_X(n\Sigma)=
\begin{cases}
0																&\text{for}~ n<0  		\\
\mc{O}_B															&\text{for}~ n=0 		\\
\mc{O}_B\oplus K_B													&\text{for}~ n=1 		\\
\mc{O}_B \oplus K_B \oplus \left(K_B^{\otimes2} \oplus K_B^{\otimes3} \oplus \ldots \oplus K_B^{\otimes n}\right)^{\oplus2} 	 			&\text{for}~ n\geq2  
\end{cases}\,,
\ee
\be
R^1\pi_*\mc{O}_X(n\Sigma)=
\begin{cases}
0 																	&\text{for}~ n>0  		\\
K_B 																	&\text{for}~ n=0  		\\
\mc{O}_B\oplus K_B														&\text{for}~ n=-1 		\\
\mc{O}_B \oplus K_B \oplus \left(K_B^{\otimes(-1)} \oplus K_B^{\otimes(-2)} \oplus \ldots \oplus K_B^{\otimes(-n+1)}\right)^{\oplus2}  	 	&\text{for}~ n\leq-2  		
\end{cases}\,.
\ee
The above results imply that $E_2^{2,0}=0$ for $n<0$ and $E_2^{0,1}=0$ for $n>0$ and, hence, that the sequence~\eqref{leray1} splits for all $n\neq 0$ . For $n=0$ this cannot be inferred in general. However, it turns out that, for our choices of base spaces and line bundles with $n=0$, one of $E_2^{2,0}$ and $E_2^{0,1}$ is always zero. This means that the sequence~\eqref{leray1} splits for all cases of interest. In conclusion, for line bundles invariant under the involution $\iota_X$, that is line bundles of the form $L={\cal O}_X(n\Sigma)\otimes\pi^*({\cal L})$ (with ${\cal L}$ a line bundle on the base), we have
\begin{equation}
H^q(X,L)=\left\{\begin{array}{lll}E_2^{q-1,1}&\mbox{ for}&n<0\\
E_2^{q-1,1}\oplus E_2^{q,0}&\mbox{ for}&n=0\\
E_2^{q,0}&\mbox{ for}&n>0\end{array}\right.\;,
\end{equation}
for $q=1,2$, where
\begin{equation}
 E_2^{i,j}=H^i(B,R^j\pi_*({\cal O}_X(n\Sigma))\otimes{\cal L})\; .
\end{equation} 
Hence, we can compute the relevant line bundle cohomologies on the CY three-fold $X$ from line bundle cohomologies on the base $B$.

\subsection{Base space choices and involutions}
\label{bases_and_involutions_maintext}
We have now expressed all relevant properties of elliptically fibered CY three-folds $X$ with two sections in terms of properties of the base manifold $B$. The final step in our model building set-up is to find a suitable explicit class of base manifolds and to develop all their required characteristics. 
Here we will present the main results with details relegated to Appendix~\ref{bases_and_involutions}.

We begin by discussing the general constraints imposed on the base manifolds. Recall from Section~\ref{fibrtwosec_maintext} that we require the existence of a freely-acting involution $\iota_X$ of $X$ which is constructed by combining a half-shift $\iota_E$ on the elliptic fibers with an involution $\iota_B$ on the base. The presence of this involution implies the existence of two sections, $\sigma$ and $\zeta$, of the fibration, as well as the factorisation of the discriminant, that is, $\Delta=\Delta_1\Delta_2^2$. This leads to two requirements on the base manifold $B$. Firstly, the three-folds $X\rightarrow B$ described by Eq.~\eqref{eq:twosectionweier} should be smooth after resolving the singularity over the locus $\Delta_2=0$. Secondly, in order for $\iota_X$ to be fixed point free, the fixed point locus of $\iota_B$ should not intersect the locus $\Delta_1=0$, where $\iota_E$ is not freely-acting on the fibers. To ensure this we require that $\iota_B$ has at most fixed points on $B$. 

The first requirement means in particular that a generic Weierstrass model \eqref{eq:generic_weierstrass} over $B$ should be smooth, which happens only when the base space $B$ is a weak Fano two-fold \cite{Morrison:2012js,Halverson:2015jua}. If the base space $B$ is not weak Fano, it is still possible to construct a corresponding smooth elliptic CY three-fold $X$ by resolving the singular Weierstrass model. After such a resolution, $X$ will have additional divisors beyond the two sections and divisors inherited from the base, which need to be taken into account when constructing and analysing models. The properties of such additional divisors depend on the choice of the base space which complicates the model building significantly. For this reason, we focus on the case of weak Fano base spaces in the present paper. More specifically, we will only consider toric weak Fano two-folds. From the 61,359 toric base spaces giving rise to flat elliptic CY three-folds \cite{Morrison:2012js}, $16$ are weak Fano and lead to generically smooth fibrations \cite{Morrison:2012js,Halverson:2015jua}. These correspond to the $16$ inequivalent reflexive polygons in two dimensions. (A list of these $16$ reflexive polytopes can be found in Figure 1 of Ref.~\cite{Braun:2013nqa}.)

\begin{figure}[h]
\begin{center}
\includegraphics[page=1,scale=1.3]{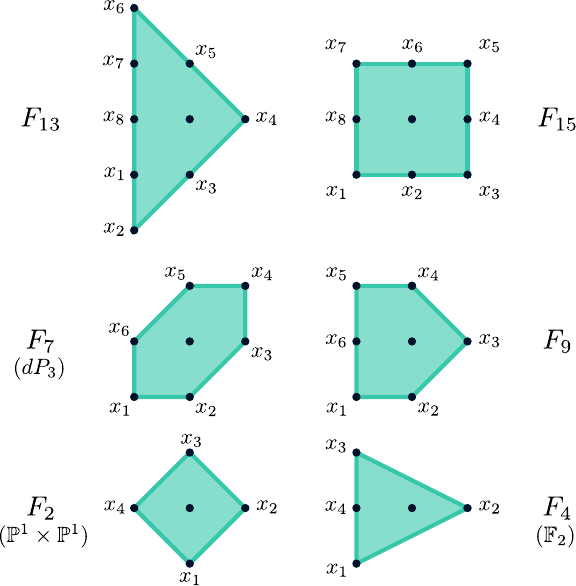}
\caption{\it The 6 reflexive polytopes corresponding to toric base spaces $B$ which lead to smooth elliptically fibered CY three-folds and allow for an involution with at most fixed points. The homogeneous coordinates $x_i$ associated to the rays are indicated and will be used throughout the paper.}
\label{fig:basespaces}
\end{center}
\end{figure}

To implement the second requirement -- the existence of an involution $\iota_B$ with at most fixed points -- we have searched these 16 surfaces for involutions which can be realised linearly on the homogeneous coordinates. We find that $6$ out of these $16$ manifolds allow for an involution of this kind which has at most fixed points. The polygons for these $6$ possible base manifolds $B$, on which we will focus for the construction of line bundle models, are shown in Fig.~\ref{fig:basespaces}.
Some basic properties of these six base spaces are listed in Table~\ref{tab:basespaces} with more details available in Appendix~\ref{bases_and_involutions}.
\begin{table}[h]
\small
\centering
\begin{tabular}{|l||l|l|l|l|l|l|}\hline
$B$&$F_2$&$F_4$&$F_7$&$F_9$&$F_{13}$&$F_{15}$\\\hline\hline
name&$\mathbb{P}^1\times\mathbb{P}^1$&$\mathbb{F}_2$&$dP_3$&--&--&--\\\hline
$h^{1,1}(B)$&$2$&$2$&$4$&$4$&$6$&$6$\\\hline
$h^{1,1}_{\rm inv}(B)$&$2$&$2$&$3$&$3$&$4$&$4$\\\hline
$c_2(B)$&$4$&$4$&$6$&$6$&$8$&$8$\\\hline
$\{{\cal C}^i\}$&$l_1=x_1$&$x_2,x_3$&$l=x_4+x_5+x_6$&$x_2,\ldots ,x_5$&$x_2,\ldots ,x_7$&$x_2,\ldots ,x_7$\\
&$l_2=x_2$&&$E_1=x_6$&&&\\
&&&$E_2=x_2$&&&\\
&&&$E_3=x_4$&&&\\\hline
$(-\lambda_i)$&$(2,2)$&$(2,0)$&$(3,-1,-1,-1)$&$(1,2,1,0)$&$(0,1,2,1,0,0)$&$(0,1,2,2,2,1)$\\\hline
$\lambda$&$8$&$8$&$6$&$6$&$4$&$4$\\\hline
$K_B$&$-2(l_1+l_2)$&$-2x_2$&$-3l+E_1$&$-x_2-x_4$&$-x_3-x_5$&$-x_3-x_7-2x_4$\\
&&&$+E_2+E_3$&$-2x_3$&$-2x_4$&$-2x_5-2x_6$\\\hline
\end{tabular}
\caption{\it Some basic properties of the six toric bases spaces $B$ which lead to smooth elliptically fibered CY three-folds and allow for an involution with at most fixed points. The coordinates $x_i$ have been defined in Fig.~\ref{fig:basespaces} and are also used to denote the curve classes defined by $x_i=0$. The row denoted $h^{1,1}_{\rm inv}(B)$ gives the dimension of the second homology invariant under the involution $\iota_B$. The first Chern class of $B$ is obtained from $c_1(B)=-K_B$ and the dual basis, $\{{\cal C}_i\}$, of curve classes can be obtained from the intersection forms $g_{ij}$ given in Appendix~\ref{bases_and_involutions}, via the relation ${\cal C}_i=g_{ij}{\cal C}^j$. Also recall the definitions $\lambda_i=K_B\cdot{\cal C}_i$ and $\lambda=\lambda^i\lambda_i$.}\label{tab:basespaces}
\end{table}

From Eq.~\eqref{isec1}, the intersection numbers of the CY three-fold $X$ can be expressed in terms of intersections on the base, that is, in terms of the quantities $\lambda_i$ and $\lambda$ given in Table~\ref{tab:basespaces} and in terms of the intersection forms $g_{ij}={\cal C}_i\cdot{\cal C}_j$ given in Appendix~\ref{bases_and_involutions}. The second Chern class of the CY three-fold $X$, which will be important to check that models are anomaly-free, can be computed from the data in Table~\ref{tab:basespaces} inserted into Eq.~\eqref{c2X}. This leads to
\begin{equation}
 c_2(X)=\left\{\begin{array}{lll}
 -4(C^0+C^{\hat{0}})+24(C^1+C^2)&\quad&B=F_2\\
 -4(C^0+C^{\hat{0}})+24C^1&\quad&B=F_4\\
12(3C^1-C^2-C^3-C^4)&\quad&B=F_7\\
12(C^1+2C^2+C^3)&\quad&B=F_9\\
 4(C^0+C^{\hat{0}})+12(C^2+2C^3+C^4)&\quad&B=F_{13}\\
 4(C^0+C^{\hat{0}})+12(C^2+2C^3+2C^4+2C^5+C^6)&\quad&B=F_{15}
 \end{array}\right. \label{C26}
 \end{equation}
The Mori cone of $X$ can be obtained from the Mori cone of the base $B$ via Eq.~\eqref{MoriX} and the generators of ${\cal M}_B$ for the six base spaces are given in Appendix~\ref{bases_and_involutions}.
 
We will also need to know the explicit action of the involution $\iota_B$ on the homogeneous coordinates $x_i$. For each of the base space choices $B$, there are multiple distinct actions on the homogeneous coordinates that give involutions with only fixed points. Most of these differ only by signs, and many of these result in the same action, $I_B$, on the curve classes in $B$ (see Eq.~\eqref{Idef}). For the purpose of this paper, it is only the action $I_B$ on the curve classes which enters the discussion, while the action on the coordinates is not explicitly used. For this reason, it is sufficient to consider one representative action $\iota_X$ for each $I_B$. These are explicitly given by
\begin{equation}
 \iota_B(x_1,\ldots ,x_{h^{1,1}(B)+2})=\left\{\begin{array}{lll}
 (x_1,x_2,-x_3,-x_4)&\quad&B=F_2\\
 (x_1,x_2,-x_3,-x_4)&\quad&B=F_4\\
 (x_4,x_5,x_6,x_1,x_2,x_3)&\quad&B=F_7\\
 (x_5,x_4,x_3,x_2,x_1,-x_6)&\quad&B=F_9\\
 (x_7,x_6,x_5,x_4,x_3,x_2,x_1,-x_8)&\quad&B=F_{13}\\
 (x_5,x_6,x_7,x_8,x_1,x_2,x_3,x_4)&\quad&B=F_{15}^{(a)}\\
 (x_7,x_6,x_5,x_4,x_3,x_2,x_1,-x_8)&\quad&B=F_{15}^{(b)}
 \end{array}\right.
 \end{equation}
Note that there are two inequivalent choices, referred to as cases $(a)$ and $(b)$, for the base space $F_{15}$. These two cases do indeed lead to different actions $I_B$ on the curve classes of $B$ and should, hence, both be taken into account. The explicit matrices $I_B$ in each case are provided in Appendix~\ref{bases_and_involutions}.

We recall from Eq.~\eqref{Dinv} that line bundles on the CY three-fold which are invariant under the involution $\iota_X$ involve curve classes on $B$ which are $\iota_B$ invariant. It is, therefore, important to have explicit expressions for all such $\iota_B$ invariant curve classes $k^i{\cal C}_i$, that is, classes which satisfy $I^i_{Bj}k^j=k^i$. Using the explicit matrices $I_B$ given in Appendix~\ref{bases_and_involutions} these invariant curves can be characterised as follows.
\begin{equation}
 \iota_B(k^i{\cal C}_i)=k^i{\cal C}_i\quad\Longleftrightarrow\quad\left\{\begin{array}{lll}
 \mbox{no constraint}&\quad&B=F_2\\
 \mbox{no constraint}&\quad&B=F_4\\
 k^1=k^2+k^3+k^4&\quad&B=F_7\\
 k^1=k^3&\quad&B=F_9\\ 
 \left\{\begin{array}{l}k^1=k^5\\k^2=k^4\end{array}\right.&\quad&B=F_{13}\\
 \left\{\begin{array}{l}k^1=k^5\\k^2=k^6\end{array}\right.&\quad&B=F_{15}^{(a)}\\
 \left\{\begin{array}{l}k^1=k^3\\k^2=k^6+2k^5-2k^3\end{array}\right.&\quad&B=F_{15}^{(b)}\\
 \end{array}\right.\label{kcons}
 \end{equation}
 Finally, we need to be able to compute the cohomology for line bundles on $B$. Methods to compute line bundle cohomology on toric spaces have been developed in Ref.~\cite{CohomOfLineBundles:Algorithm,cohomCalg:Implementation} and we will use the accompanying code \cohomCalg for our calculations. 

\section{Line bundle models}
\label{line_bundle_models}
All mathematical ingredients for the construction of heterotic line bundle models on elliptically fibered CY three-folds with two sections are now in place. In this section, we review the general construction of line bundle models and the structure of their low-energy spectrum (see Refs.~\cite{Anderson:2012yf} for a comprehensive account) as well as some particular features of line bundle models for elliptically fibered CY three-folds with two sections. A systematic line bundle model search on these CY manifolds will be presented in the next section.

\subsection{Construction of line bundle models}
A heterotic $E_8\times E_8$ line bundle model is defined by two ingredients: a CY three-fold $X$ and a line bundle sum
\begin{equation}
 V=\bigoplus_{a=1}^{r}L_a\; , \label{Vdef}
\end{equation}
with rank $r={\rm rk}(V)$, where $L_a$ are line bundles on $X$. We think of $V$ as the bundle in the ``observable" $E_8$ sector and will comment on the fate of the hidden sector below. In practice, it is useful to represent line bundles relative to an integral basis $\{D_I\}$ of divisor classes on $X$ and we write $L={\cal O}_X({\bf k})$ for a line bundle $L$ with first Chern class $c_1(L)=k^ID_I$. Using this notation, the line bundles 
\begin{equation} 
 L_a={\cal O}_X({\bf k}_a)\; ,
\end{equation} 
can each be represented by an integer vector ${\bf k}_a$ and the entire line bundle sum $V$ by a $h^{1,1}(X)\times r$ integer matrix $K=({k^I}_a)$.

The data $(X,V)$ is subject to three conditions which are required for the consistency of the model. Firstly, we need to be able to embed the structure group of the bundle $V$ into $E_8$. Apart from the obvious rank constraint , $r\leq 8$, this requires the vanishing for the first Chern class of $V$, that is,
\begin{equation}
 c_1(V)=\sum_{a=1}^{r}c_1(L_a)\stackrel{!}{=}0\qquad\Longleftrightarrow\qquad \sum_{a=1}^r{\bf k}_a\stackrel{!}{=}0\; . \label{c10}
\end{equation} 
Provided this is satisfied the structure group of $V$ is generically $S(U(1)^r)$ which allows for an embedding into $E_8$ via the sub-group chain $S(U(1)^r)\subset SU(r)\subset E_8$. 

The second constraint originates from the requirement that the bundle $V$ preserve supersymmetry. It can be formulated in terms of the slope of a line bundle $L={\cal O}_X({\bf k})$ which is defined as
\begin{equation}
 \mu_X(L)=\int_XJ\wedge J\wedge c_1(L)=d_{IJK}t^It^Jk^K\; ,
\end{equation}
where $d_{IJK}=D_I\cdot D_J\cdot D_K$ are the triple intersection numbers of $X$. Supersymmetry of $V$ requires that the slope of all line bundles $L_a$ vanishes simultaneously somewhere in the interior, $\mathring{\cal K}_X$, of the K\"ahler cone of $X$. This means that the equations
\begin{equation}
 \mu_X(L_a)=d_{IJK}t^It^Jk_a^K\stackrel{!}{=}0\; , \label{slope0}
\end{equation} 
for $a=1,\ldots , r$ should have a common solution ${\bf t}\in\mathring{\cal K}_X$. 

The third requirement is anomaly cancellation which demands the existence of a hidden bundle $\tilde{V}$ (which should also be supersymmetric) and a holomorphic curve $W\subset X$ (around which a five-brane wraps) such that
\begin{equation}
 {\rm ch}_2(V)+{\rm ch}_2(\tilde{V})-{\rm ch}_2(TX)=[W]\; .\label{anomcond}
\end{equation} 
A practical way to ensure that this condition can be satisfied is to demand that
\begin{equation}
 c_2(TX)-c_2(V)\in{\cal M}_X\; , \label{anomcond1}
\end{equation}
where ${\cal M}_X$ is the Mori cone (that is, the cone of effective curve classes) on $X$. Indeed, provided Eq.~\eqref{anomcond1} holds, we can always choose a holomorphic curve $W$ with $[W]= c_2(TX)-c_2(V)$, so that the anomaly condition~\eqref{anomcond} is satisfied for a trivial hidden bundle (although different choices for $[W]$ combined with a non-trivial hidden bundle $\tilde{V}$ are usually possible as well). To check the condition~\eqref{anomcond1} explicitly, we introduce a basis $\{C^I\}$ of curve classes on $X$, dual to our basis $\{D_I\}$ of divisor classes such that $C^I\cdot D_J=\delta^I_J$. Then, the second Chern class of $V$ can be written as
\begin{equation}
 c_2(V)=-\frac{1}{2}d_{IJK}\left(\sum_{a=1}^rk_a^Ik_a^J\right)C^K\; ,
\end{equation} 
and can be compared with the second Chern class of the tangent bundle expressed in the same basis as $c_2(TX)=c_{2I}(TX)C^I$. 

Provided the above three conditions are satisfied we have a consistent line bundle model with defining data $(X,V)$. For phenomenological purposes, we would like to quotient this model by a freely-acting symmetry, $\Gamma$, of the CY three-fold $X$ and obtain a model on the quotient CY manifold $\hat{X}=X/\Gamma$. The symmetry $\Gamma$ should lift to the bundle $V$ (in mathematical terminology, the bundle $V$ should have a $\Gamma$-equivariant structure) so that it descends to a bundle $\hat{V}\rightarrow\hat{X}$ on the quotient. The final step is to add a Wilson line bundle ${\cal W}$ so that the complete ``downstairs" bundle is $\hat{V}\oplus {\cal W}$. 

For the purpose of this paper, we would like to carry out a necessary (but not quite sufficient) check, adapted to our algorithmic model search, for equivariance. We will check that the line bundle sum $V$ in Eq.~\eqref{Vdef} is $\Gamma$-invariant, which is the case iff $\gamma^*(V)\cong V$ for all $\gamma\in\Gamma$. A line bundle sum is $\Gamma$-invariant iff $\gamma^*(V)$ amounts to a permutations of the various line bundles. As a further simplification, we focus on cases where these permutations are trivial, so that every line bundle $L_a$ is invariant by itself. Given that the symmetry group we consider is $\Gamma=\mathbb{Z}_2$, generated by an involution $\iota_X$, we therefore demand that 
\begin{equation}
 \iota_X^*(c_1(L_a))=c_1(L_a)\; ,
\end{equation}
for all $a=1,\ldots ,r$.

\subsection{Spectrum of line bundle models}\label{spectrum}
The (observable) low-energy gauge group is the commutant of $S(U(1)^r)$, the structure group of the line bundle sum $V$, within $E_8$. From a phenomenological point of view, the most attractive choice is $r=5$ and in this case the low-energy gauge group is given by
\begin{equation}
 G=SU(5)\times S(U(1)^5)\; .
\end{equation} 
Under the maximal sub-group $SU(5)\times SU(5)\subset E_8$ the adjoint representation of $E_8$ branches as
\begin{equation}
{\bf 248}_{E_8}\rightarrow ({\bf 24},{\bf 1})\oplus({\bf 1},{\bf 24})\oplus  ({\bf 10},{\bf 5})\oplus (\overline{\bf 10},\bar{\bf 5})\oplus (\bar{\bf 5},{\bf 10})\oplus ({\bf 5},\overline{\bf 10})\; .
\end{equation}
Embedding $G\subset SU(5)\times SU(5)$ the representations on the right-hand side branch further into the $G$-representations
\begin{equation}
 \begin{array}{lllllll}
  ({\bf 24},{\bf 1})&\rightarrow& {\bf 24}_{\bf 0}&,&({\bf 1},{\bf 24})&\rightarrow& \bigoplus_{a,b} {\bf 1}_{{\bf e}_a-{\bf e}_b}\\
  ({\bf 10},{\bf 5})&\rightarrow&\bigoplus_a{\bf 10}_{{\bf e}_a}&,& (\overline{\bf 10},\bar{\bf 5})&\rightarrow&\bigoplus_{a} {\overline{\bf 10}}_{-{\bf e}_a}\\
  (\bar{\bf 5},{\bf 10})&\rightarrow&\bigoplus_{a<b}\bar{\bf 5}_{{\bf e}_a+{\bf e}_b}&,&({\bf 5},\overline{\bf 10})&\rightarrow&\bigoplus_{a<b}{\bf 5}_{-{\bf e}_a-{\bf e}_b}\; .
 \end{array}\label{multiplets}
\end{equation}
Here, ${\bf r}_{\bf q}$ denotes an $SU(5)$ representation ${\bf r}$ with $S(U(1)^5)$ charge ${\bf q}$, a five-dimensional integer vector, defined up to multiples of $(1,1,1,1,1)$. Further, ${\bf e}_a$, where $a=1,\ldots ,5$, denote the five-dimensional standard unit vector. This means that, for example, the multiplet ${\bf 10}_{{\bf e}_a}$ carries charge $+1$ under the $a^{\rm th}$ $U(1)$ symmetry in $S(U(1)^5)$ and is uncharged under the others. For simplicity, we will frequently replace the sub-scipt ${\bf e}_a$ by $a$, so that, for example, ${\bf 10}_{{\bf e}_a}$ is written as ${\bf 10}_a$.

 The $G$-representations on the right-hand sides of Eq.~\eqref{multiplets} provide a list of possible multiplets which can arise in the effective theory. If we formally assign the charge ${\bf e}_a$ to the line bundle $L_a$ then every tensor product of these line bundles and their duals acquires an induced charge, in the obvious way. For example, the line bundles $L_a\otimes L_b$ then carry charges ${\bf e}_a+{\bf e}_b$. Then, every multiplet ${\bf r}_{\bf q}$ in \eqref{multiplets} can be associated to the line bundle, $L\in\{ L_a, L_a\otimes L_b, L_a\otimes L_b^*, L_a^*\otimes L_b^*\}$, with the same $S(U(1)^5)$ charge ${\bf q}$. The first cohomology, $h^1(X,L)$, of this associated line bundle counts the multiplicity of a multiplet. The details of this correspondence are provided in Table~\ref{tab:coh}.
 \begin{table}[h]
 \begin{center}
 \begin{tabular}{|c|c|c|c|c|}\hline
 multiplet&$S(U(1)^5)$ charge&associated $L$&contained in\\\hline\hline
 ${\bf 10}_a$&${\bf e}_a$&$L_a$&$V$\\\hline
 $\overline{\bf 10}_a$&$-{\bf e}_a$&$L_a^*$&$V^*$\\\hline
 $\bar{\bf 5}_{a,b}$&${\bf e}_a+{\bf e}_b$&$L_a\otimes L_b$&$\wedge^2 V$\\\hline
 ${\bf 5}_{a,b}$&$-{\bf e}_a-{\bf e}_b$&$L_a^*\otimes L_b^*$&$\wedge^2 V^*$\\\hline
 ${\bf 1}_{a,b}$&${\bf e}_a-{\bf e}_b$&$L_a\otimes L_b^*$&$V\otimes V^*$\\\hline
 \end{tabular}
 \caption{\it A list of $SU(5)\times S(U(1)^5)$ multiplets in the low-energy theory and their associated line bundles. The multiplicity of each multiplet is computed by the cohomology $h^1(X,L)$ of the associated line bundle $L$.}\label{tab:coh}
 \end{center}
 \end{table}
We recall that all line bundles $L_a$ (and, hence, their duals and all their tensor products) need to have a vanishing slope which implies that $h^0(X,L)=h^3(X,L)=0$ for all line bundles $L$ which appear in Table~\ref{tab:coh} (with the exception of $L_a\otimes L_a^*={\cal O}_X$). This means that the chiral asymmetry of multiplets is computed by the index, that is,
\begin{eqnarray}
 \#{\bf 10}_a-\#\overline{\bf 10}_a&=&h^1(X,L_a)-h^1(X,L_a^*)=-{\rm ind}(L_a)\\
 \#\bar{\bf 5}_{a,b}-\#{\bf 5}_{a,b}&=&h^1(X,L_a\otimes L_b)-h^1(X,L_a^*\otimes L_b^*)=-{\rm ind}(L_a\otimes L_b)\; .
\end{eqnarray} 
 
What should be required of this spectrum for a physically promising model? First, we remark that the additional gauge bosons associated to the $S(U(1)^5)$ symmetry do not constitute a phenomenological problem. They are either heavy due to the Green-Schwarz effect or can acquire mass due to spontaneous symmetry breaking induced by the $SU(5)$ singlet fields in the last row of Table~\ref{tab:coh}. For the correct total chiral asymmetry in the ${\bf 10}$--$\overline{\bf 10}$ sector we should require that
\begin{equation}
 {\rm ind}(V)=\sum_a{\rm ind}(L_a)\stackrel{!}{=}-3|\Gamma|\; . \label{indcond}
\end{equation}
This condition guarantees three chiral ${\bf 10}$ families after dividing by the symmetry $\Gamma$ with order $|\Gamma|$. Fortunately, for $SU(5)$ bundles $V$, we have ${\rm ind}(V)={\rm ind}(\wedge^2 V)$ so that the above condition also guarantees the correct chiral asymmetry in the $\bar{\bf 5}$--${\bf 5}$ sector.

For models where each line bundle $L_a$ is equivariant by itself we should also ensure that the chiral asymmetry in every $S(U(1)^5)$ charge sector has the ``correct" sign, that is, that there is no excess of $\overline{\bf 10}_a$ over ${\bf 10}_a$ multiplets for any $a$ and no excess of ${\bf 5}_{a,b}$ over $\bar{\bf 5}_{a,b}$ multiplets for any $a,b$. To avoid such "wrong" chiral asymmetries we impose that
\begin{equation}
 {\rm ind}(L_a)\leq 0\;,\qquad {\rm ind}(L_a\otimes L_b)\leq 0 \label{chiralcond}
\end{equation}
for all $a,b=1,\ldots ,5$ with $a<b$.  

Further, for a ``clean" spectrum we should demand the absence of ${\bf 10}$--$\overline{\bf 10}$ vector-like pairs and the presence of precisely one $\bar{\bf 5}$--${\bf 5}$ vector-like pair to account for the Higgs doublets. These requirements can be expressed as
\begin{equation}
 h^1(X,V^*)=\sum_a h^1(X,L_a^*)\stackrel{!}{=}0\;,\qquad h^1(X,\wedge^2V^*)=\sum_{a<b}h^1(X,L_a^*\otimes L_b^*)\stackrel{!}{=}1\; . \label{cleancond}
\end{equation} 
If all these conditions are satisfied, a model with the precise MSSM spectrum can usually be obtained after dividing by the symmetry $\Gamma$. The only exotic states left in the upstairs theory are the Higgs triplets contained in the $\bar{\bf 5}$--${\bf 5}$ vector-like pair. They can normally be projected out for a suitable choice of equivariant structure and Wilson line, while the Higgs doublets can be kept. 

While the conditions~\eqref{cleancond} give rise to a clean standard model spectrum in this way they are by no means necessary. Additional vector-like states as they arise when the conditions~\eqref{cleancond}  are violated can receive a mass from superpotential terms of the form ${\bf 1}\,{\bf 10}\,\overline{\bf 10}$ and ${\bf 1}\,{\bf 5}\,\bar{\bf 5}$ (or even higher-dimensional operators with multiple singlet insertions), when the $SU(5)$ singlet fields acquire a vacuum expectation value. It is worth noting that these singlet fields, which appear in the last row of Table~\ref{tab:coh}, carry a $S(U(1)^5)$ charge so that the presence of such terms is constrained by the $S(U(1)^5)$ symmetry. Switching on singlet vacuum expectation values (which, in the language of the effective theory, has to be done preserving D- and F-flatness) corresponds to moving in the bundle moduli space and deforming the line bundle sum $V$ to a bundle with non-Abelian structure group. It is perfectly possible and likely to happen in many cases, that  unwanted vector-like states can be removed in this way, but checking this is a matter of detailed analysis within each model. 

\subsection{Line bundle models and elliptic fibrations}
We would like to discuss a number of specific features which arise for line bundle models on elliptically fibered CY three-folds with a freely-acting involution of the kind considered in this paper.

First, recall that on such CY three-folds, the line bundles $L_a$ can be written as $L_a={\cal O}_X(k_a^ID_I)$, where the basis $\{D_I\}=\{D_0=\sigma(B),D_{\hat{0}}=\zeta(B),D_i=\pi^{-1}({\cal C}_i)\}$ of divisor classes has been defined in Eq.~\eqref{Dbasis}. Hence, every line bundle can be represented by an integer vector
\begin{equation}
 {\bf k}_a=({k^I}_a)=({k^0}_a,{k^{\hat{0}}}_a,{k^i}_a)^T\; .
\end{equation}
From Eq.~\eqref{lbinv}, $\iota_X$ invariance of the line bundles means that
\begin{equation}
 {k^0}_a={k^{\hat{0}}}_a\;\mbox{ and }\;{k^i}_a\mbox{ of the form \eqref{kcons}}\mbox{, for all }a=1,\ldots ,5\; . \label{kcons1}
\end{equation} 
Alternatively, we will also represent the entire line bundle sum $V$ by an $h^{1,1}(X)\times 5$ integer matrix
\begin{equation}
  K=({k^I}_a)
\end{equation}
with the columns corresponding to the line bundles. In view of the condition $c_1(V)\sim\sum_a{\bf k}_a\stackrel{!}{=}0$ and Eq.~\eqref{kcons1}, the number of independent integers in $K$ is given by $4(h^{1,1}_{\rm inv}(B)+1)$, where the numbers $h^{1,1}_{\rm inv}(B)$ have been listed in Table~\ref{tab:basespaces}.

It is interesting to consider the total chiral asymmetry of such models which, from Eqs.~\eqref{indL} and \eqref{c10}, is given by
\begin{equation}
 {\rm ind}(V)=\sum_a{\rm ind}(L_a)=-\frac{1}{6}d_{IJK}\sum_a k_a^Ik_a^Jk_a^K\; . \label{indV}
\end{equation} 
Since the pure base intersection numbers $d_{ijk}$ vanish from Eq.~\eqref{isec1} this means not all of the integers $k^0_a$ (and $k^{\h{0}}_a$) can be zero if we want to obtain a chiral model. In other words, some of the line bundles $L_a$ need to have a first Chern class with a non-zero coefficient in the direction of $\sigma(B)$ (and $\zeta(B)$), leading to non-trivial bundle upon restriction to the fibers, in order to generate a non-vanishing chiral asymmetry. Heterotic F-theory duality in its current formulation requires bundles which are flat on the fibers \cite{Friedman:1997yq}. Hence, at present, there is no obvious F-theory dual for chiral line bundle models. It would be interesting to study if heterotic F-theory duality can be extended to such cases.

Finally, to calculate line bundle cohomology, we use the results of Section~\ref{sec:lbs} to express the cohomology of line bundles on $X$ in terms of line bundle cohomology on the toric two-fold base $B$. The latter is then computed using the code for line bundle cohomology on toric spaces developed in Ref.~\cite{CohomOfLineBundles:Algorithm,cohomCalg:Implementation}. 

The CY manifolds used in this paper can also be realised as hypersurfaces in toric four-folds, as described in Appendix~\ref{app:toric2sectionmodels} and this provides us with an alternative method to calculate line bundle cohomology. More specifically, a line bundle ${\cal L}$ on the toric ambient four-fold ${\cal A}$ and its restriction $L$ to the CY three-fold $X$ are related by the Koszul sequence
\begin{equation}
 0\rightarrow N^*\otimes {\cal L}\rightarrow{\cal L}\rightarrow L\rightarrow 0\; , \label{koszul}
\end{equation}
where $N$ is the anti-canonical bundle of ${\cal A}$. The cohomologies of ${\cal L}$ and $N^*\otimes{\cal L}$  can again be computed  using the code in Ref.~\cite{CohomOfLineBundles:Algorithm,cohomCalg:Implementation}, this time applied to the ambient toric four-fold ${\cal A}$. The cohomology of $L$ can then be inferred from the long exact sequence associated to the above Koszul sequence. We have carried this out for $\iota_X$ invariant line bundles and we find that the cohomology of $L$ can be determined easily from the long exact sequence without the need to compute ranks of maps (that is, the sequence always splits). Moreover, the results always agree with the previous method, based on the Leray spectral sequence. 

\section{Systematic model search}\label{sec_scan}
We have now collected all ingredients  required for the construction of heterotic line bundle models on elliptically fibered CY three-folds with a freely-acting involution. In this section, we describe the results of a systematic scan, searching for physically promising models, which covers the six possible base manifolds in Table~\ref{tab:basespaces} and rank five line bundles sums. We also illustrate our results by explicitly presenting a specific model found in this scan.

\subsection{Scan results}\label{scan_results}
Our search has been carried out for all six base spaces in Table~\ref{tab:basespaces}.  For each base space, we have scanned over all line bundle sums $K=({k^I}_a)$ which satisfy Eq.~\eqref{kcons1} (so that each constituent line bundle is $\iota_X$ invariant) and whose entries are bounded by $|{k^I}_a|\leq k_{\rm max}$. The quantity $k_{\rm max}$ has been maximised in view of computational limitations on a desktop machine and its values are listed in Table~\ref{tab:scan}.

From the line bundle sums generated in this way, we have selected the physically promising ones by the following set of criteria.
\begin{itemize}
 \item Eq.~\eqref{c10}, $c_1(V)=0$, is satisfied so that the structure group is $S(U(1)^5)$.
 \item There is a locus in K\"ahler moduli space where the slopes of all line bundles $L_a$ vanish, that is, Eq.~\eqref{slope0} is satisfied for all $L_a$. This means the line bundle sum $V$ preserves supersymmetry. 
 \item The anomaly condition~\eqref{anomcond1} is satisfied.
 \item Following Eq.~\eqref{indcond}, the index of the line bundle sum satisfies ${\rm ind}(V)=-6$. This guarantees three chiral families of quarks and leptons after taking the quotient by the involution $\iota_X$. 
 \item The indices of $L_a$ and $L_a\otimes L_b$ are constrained by Eqs.~\eqref{chiralcond} to avoid a chiral asymmetry with the wrong sign in any $S(U(1)^5)$ charge sector. 
\end{itemize} 
The number of models satisfying these conditions is given in the last column in Table~\ref{tab:scan}.
\begin{table}[h]
\begin{center}
\begin{tabular}{|c|c|c|c|}
\hline
base $B$&$k_{\rm max}$&$k_{\rm mod}$& $\#$models \\ \hline \hline
$F_2$&$10$&--&$0$\\\hline
$F_4$&$10$&--&$0$\\\hline
$F_7$ &$10$&$4$& $54$ \\ \hline
$F_9$ &$7$&$6$&$22$ \\ \hline
$F_{13}$ &$3$&$3$&$\geq46$ \\ \hline
$F_{15}^{(a)}$ &$3$&$3$&$\geq236$ \\ \hline
$F_{15}^{(b)}$ &$3$&$3$&$\geq84$ \\ \hline\hline
total&--&--&$\geq 442$\\\hline
\end{tabular}
\caption{\it The number of phenomenologically interesting models found for each of the six base manifolds. The scan was carried out over all line bundle models with $|{k^I}_a|\leq k_{\rm max}$ and $k_{\rm mod}$ gives the largest value of $|{k^I}_a|$ which arises in a physically interesting model.}\label{tab:scan}
\end{center}
\end{table}
The list of all integer matrices $K$ for those physically promising models can be downloaded from Ref.~\cite{bundledata}.

The number of models found for each base manifold $B$ can be qualitatively understood by considering the number $h^{1,1}_{\rm inv}(B)+1$ (see Table~\ref{tab:basespaces} for the values of $h^{1,1}_{\rm inv}(B)$) of independent integers which specify a $\iota_X$ invariant line bundle. For $B=F_2,F_4$ we have $h^{1,1}_{\rm inv}(B)=2$ and this does evidently not provide enough freedom to allow for interesting models. For $B=F_7, F_9$ we have $h^{1,1}_{\rm inv}(B)=3$ and in these case we are able to find all physically promising models by extending the scan to a sufficiently large $k_{\rm max}$. For the last two base spaces, $B=F_{13},F_{15}$, with $h^{1,1}_{\rm inv}(B)=4$ the space of line bundle sums becomes quite large and we have only carried out a partial scan for $k_{\rm max}=3$. For those cases, the number of interesting models exceeds the numbers given in Table~\eqref{tab:scan}. In total, we find $442$ models for all six base manifolds. 

\begin{figure}[h]
\begin{center}
\includegraphics[page=1,scale=.4]{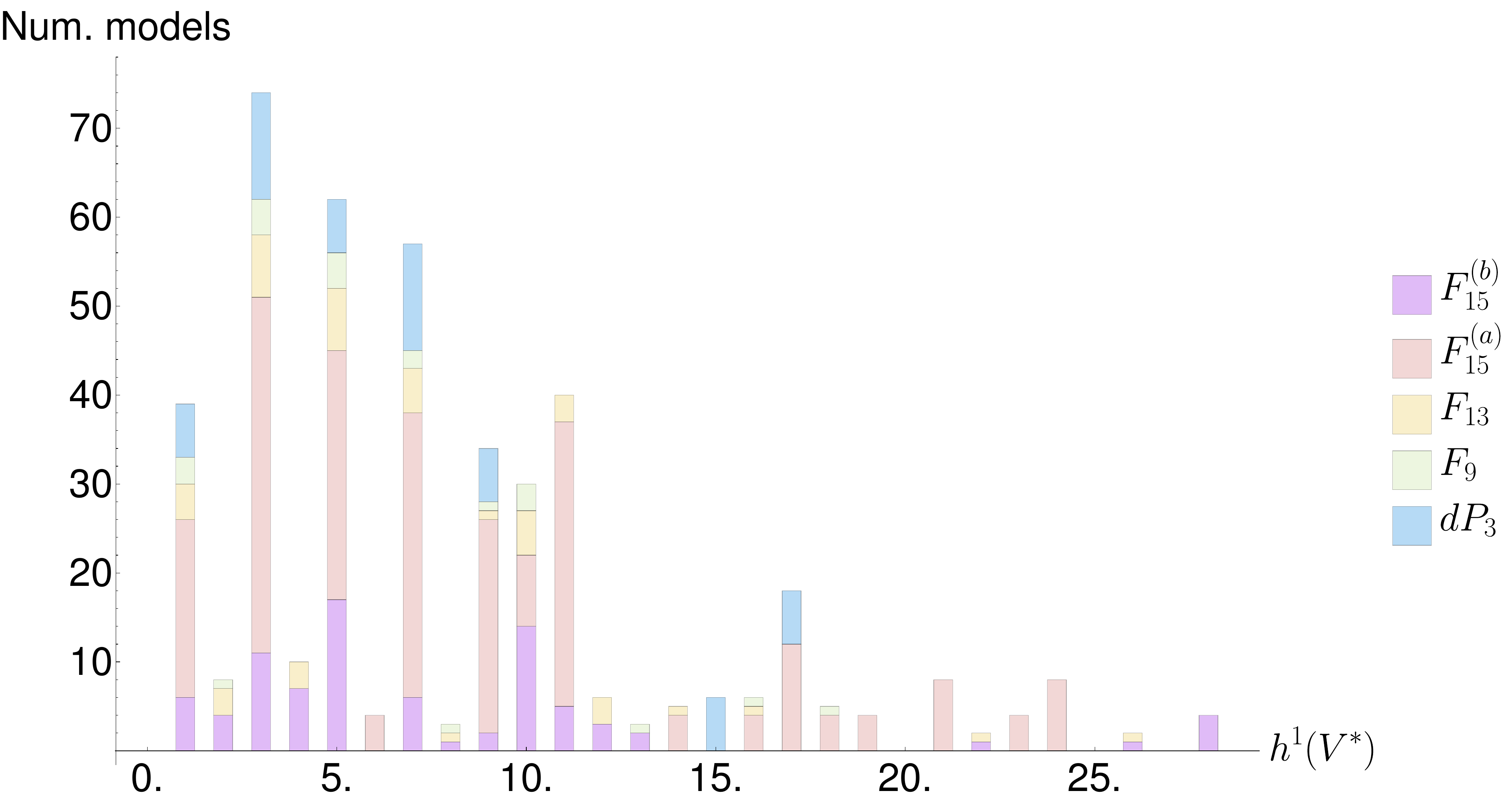}
\caption{Frequency plot of $h^1(V^*)$ which gives the number of $\overline{\bf 10}$ multiplets, combined for all base spaces.}\label{fig:10bar}
\end{center}
\end{figure}

\begin{figure}[h]
\begin{center}
\includegraphics[page=1,scale=.4]{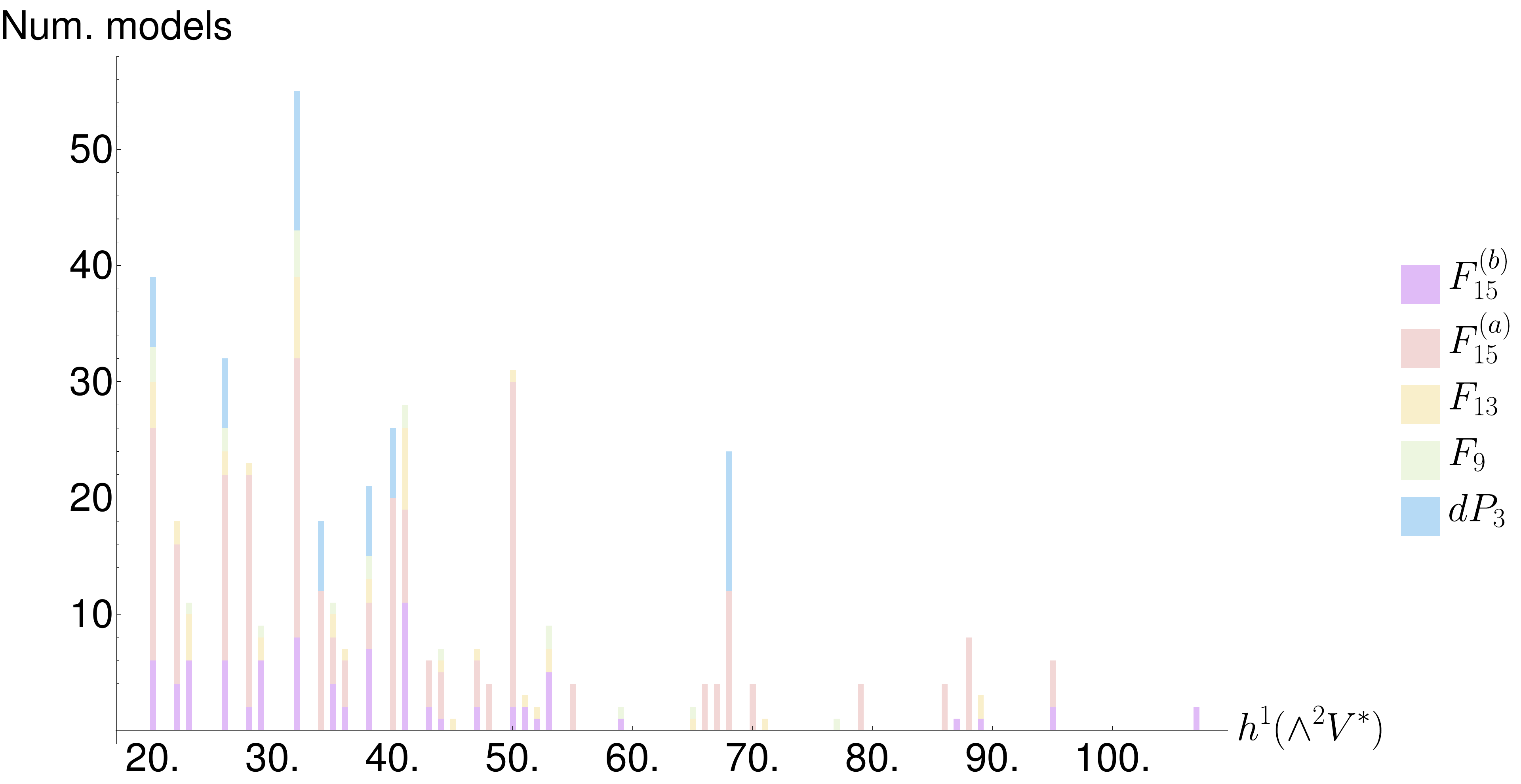}
\caption{Frequency plot of $h^1(\wedge^2V^*)$ which gives the number of ${\bf 5}$ multiplets, combined all base space spaces.}\label{fig:5}
\end{center}
\end{figure}

\begin{figure}[h]
\begin{center}
\includegraphics[page=1,scale=.4]{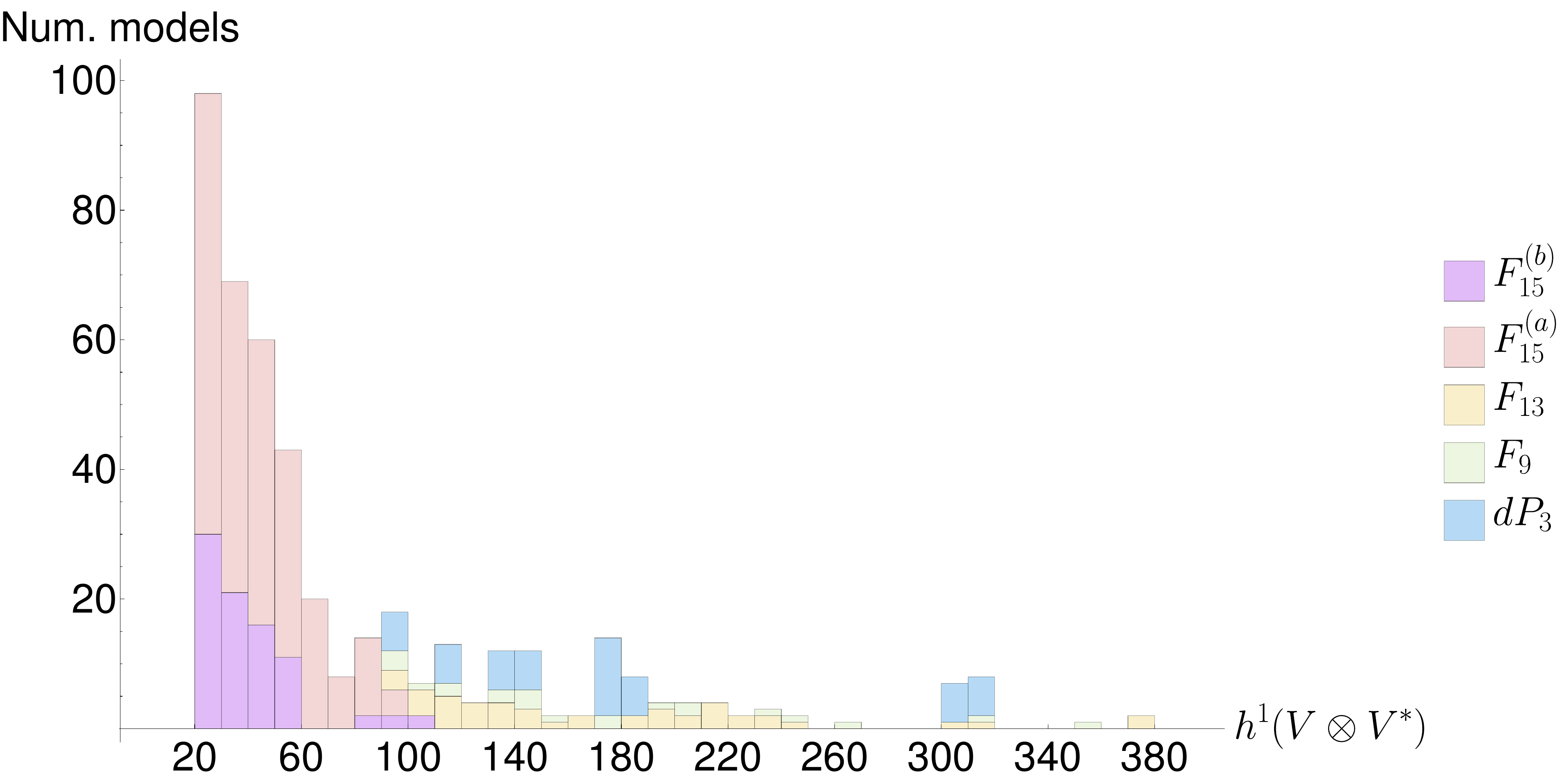}
\caption{Frequency plot of $h^1(V \otimes V^*)$ which gives the number of singlet fields, combined for all base spaces.}\label{fig:singlets}
\end{center}
\end{figure}

For those $442$ models, we have also determined the complete spectrum by computing all relevant line bundle cohomologies, as explained in Section~\ref{spectrum}. The results of these computations are summarised in Figs.~\ref{fig:10bar}, \ref{fig:5} and \ref{fig:singlets} which provide frequency plots for the number of $\overline{\bf 10}$ multiplets, ${\bf 5}$ multiplets and singlets, respectively.
As these plots show, there is unfortunately no model without additional vector-like pairs. From Fig.~\ref{fig:10bar}, there exists always at least one ${\bf 10}$--$\overline{\bf 10}$ vector-like pair and frequently many more. Fig.~\ref{fig:5} shows that the situation is worse for ${\bf 5}$--$\bar{\bf 5}$ vector-like pairs, where the minimal number is $20$. This large number of vector-like pairs comes as a surprise, given the experience with line bundle models on complete intersection CY three-folds~\cite{Anderson:2011ns,Anderson:2012yf,Anderson:2013xka} where imposing the correct chiral asymmetry frequently resulted in the absence of additional vector-like states. 

As we have argued above, these vector-like states do not necessarily render the models unphysical since they can be given a mass via couplings to singlet fields with non-vanishing vacuum expectation values. Fig.~\ref{fig:singlets} shows that our models do indeed have a significant number of such singlet fields and it is likely that they can be used to remove unwanted vector-like pairs in many cases. Analysing this is a matter of more detailed model building which is beyond the scope of the present paper. 

\subsection{An example model}
As an illustration, we will now present one of the physically interesting models from the previous sub-section in detail. Our example is for the base space $B=F_7=dP_3$ which, following Appendix~\ref{using_delpezzo_appendix}, has a basis of curve classes $\{{\cal C}^i\}=\{l,E_1,E_2,E_3\}$ with dual basis $\{{\cal C}_i\}=\{l,-E_1,-E_2,-E_3\}$. Line bundles will be represented relative to the basis $\{D_I\}=\{D_0,D_{\hat{0}},D_i\}$, where
\begin{equation}
D_0=\sigma(B)\;,\quad D_{\hat{0}}=\zeta(B)\;,\quad D_i=\pi^{-1}({\cal C}_i)\;,\quad i=1,\ldots ,4\; .
\end{equation} 
Writing $L_a={\cal O}_X({k^I}_a D_I)$ as before, the integer matrix $K=({k^I}_a)$ which defines our example is given by
\begin{equation}
 K=
\left(\begin{array}{rrrrr}
   -1&0&0&0&1\\
   -1&0&0&0&1\\
   1&0&0&0&-1\\
   -1&1&1&1&-2\\
   1&-1&-1&-1&2\\
   1&0&0&0&-1
\end{array}\right)\; ,\label{Kex}
\end{equation}   
with every column representing one of the line bundles $L_a$. First, we note that the columns sum up to zero so that the constraint~\eqref{c10}, $c_1(V)\sim\sum_a{\bf k}_a=0$, is indeed satisfied. Further, we see that the matrix is consistent with $\iota_X$ invariance of each line bundle. Specifically, the first and second rows are identical, so that ${k^0}_a={k^{\hat{0}}}_a$, and the last four rows satisfy ${k^1}_a={k^2}_a+{k^3}_a+{k^4}_a$, in accordance with Eq.~\eqref{kcons1}.

Following the list of required properties in Section~\ref{scan_results}, we should next check that the slope of all line bundles vanishes somewhere in the K\"ahler cone. Given the structure of the matrix~\eqref{Kex} it is sufficient to do this for the line bundles $L_1$ and $L_2$ whose slopes are explicitly given by
\bea
\mu_X(L_1) &=&-8 t_0^2 + 8 t_0 t_1 - 2 t_1^2 + 2 t_2^2 - 4 t_0 t_3 + 2 t_3^2  \nonumber\\
&& - 4 t_0 t_4 + 2 t_4^2 + 8 t_1 t_{\hat{0}} - 4 t_3 t_{\hat{0}} - 4 t_4 t_{\hat{0}} - 8 t_{\hat{0}}^2  \; , \\ 
\mu_X(L_2) &=&-2 t_0 t_2 + 2 t_0 t_3 - 2 t_2 t_{\hat{0}} + 2 t_3 t_{\hat{0}} \; .
\eea
It can be verified that $\mu_X(L_1)=\mu_X(L_2)=0$ for 
\begin{equation}
t_0=\frac{5}{8}\;,\quad t_{\hat{0}}=\frac{5}{8}\;,\quad t_1=\frac{43}{12}\;,\quad t_2=1\;,\quad t_3=1\;,\quad t_4=\frac{19}{12}  \; .
\end{equation}
Comparison with Eq.~\eqref{MCF7} shows that this point it indeed in the interior of the K\"ahler cone of $X$.

Next, we should verify the anomaly condition for this model. The second Chern class of the bundle $V$ is given by
\be
c_2(V)=18(F-N)-2N+10\,\sigma(l)-4\,\sigma(E_1)-6\,\sigma(E_3)\,
\ee
and comparing this with the second Chern class of the tangent bundle~\eqref{C26} gives
\be
c_2(X)-c_2(V)=54(F-N)+2N+8\sigma(l-E_1)+12\sigma(l-E_2)+6\sigma(l-E_3)\; .
\ee
Since $F-N$ and $N$ are effective curves and, from Appendix~\ref{using_delpezzo_appendix}, $l-E_a$ are effective curves in $dP_3$ this class is indeed effective. Hence, the anomaly can be cancelled by wrapping a five-brane on a holomorphic curve in this class. Finally, using Eqs.~\eqref{indL} and \eqref{indV} we can verify that indeed ${\rm ind}(V)=-6$, ${\rm ind}(L_a)\leq 0$ and ${\rm ind}(L_a\otimes L_b)\leq 0$ for all $a,b=1,\ldots ,5$.

In summary, we have seen that this line bundle sum is invariant under the involution and provides a consistent model with the correct chiral asymmetry.

To determine the spectrum in more detail we consider line bundle cohomologies. For the line bundle sum $V$ we find
\begin{equation}
 h^\bullet(X,V)=(0,7,1,0)\; .
\end{equation}
This overall result originates from the individual line bundle cohomologies
\begin{equation}
 h^\bullet(X,L_1)=(0,1,1,0)\;,\quad h^\bullet(X,L_5)=(0,6,0,0)\; ,
\end{equation} 
with the cohomologies of all other $L_a$ vanishing. For $\wedge^2 V$ we have
\begin{equation}
 h^\bullet(X,\wedge^2 V)=(0,26,20,0)\; ,
\end{equation}
obtained as a sum of the cohomologies
\begin{align}
 &h^\bullet(X,L_1\otimes L_2)=h^\bullet(X,L_1\otimes L_3)=h^\bullet(X,L_1\otimes L_4)=(0,2,0,0)\\
 &h^\bullet(X,L_2\otimes L_3)=h^\bullet(X,L_2\otimes L_4)=h^\bullet(X,L_3\otimes L_4)=(0,3,3,0)\\
 &h^\bullet(X,L_2\otimes L_5)=h^\bullet(X,L_3\otimes L_5)=h^\bullet(X,L_4\otimes L_5)=(0,1,1,0)\\
 &h^\bullet(X,L_1\otimes L_5)=(0,8,8,0)\; . 
\end{align}
Combining these results the chiral spectrum of the model is
\begin{equation}
 6\,{\bf 10}_5\;,\; 2\,\bar{\bf 5}_{1,2}\;,\; 2\,\bar{\bf 5}_{1,3}\;,\; 2\,\bar{\bf 5}_{1,4}\; ,
\end{equation}
and we have the additional vector-like pairs
\begin{align} 
 &({\bf 10}_1\oplus\overline{\bf 10}_1)\;,\;3\,({\bf 5}_{2,3}\oplus\bar{\bf 5}_{2,3})\;,\;3\,({\bf 5}_{2,4}\oplus\bar{\bf 5}_{2,4})\;,\;3\,({\bf 5}_{3,4}\oplus\bar{\bf 5}_{3,4})\\
 &8\, ({\bf 5}_{1,5}\oplus\bar{\bf 5}_{1,5})\;,\; ({\bf 5}_{2,5}\oplus\bar{\bf 5}_{2,5})\;,\; ({\bf 5}_{3,5}\oplus\bar{\bf 5}_{3,5})\;,\; ({\bf 5}_{4,5}\oplus\bar{\bf 5}_{4,5})\; .
\end{align}
This illustrates the aforementioned proliferation of vector-like pairs. In addition, across all $S(U(1)^5)$ charge sectors, the model has $h^1(V\otimes V^*)=94$ singlet fields whose $S(U(1)^5)$ charges we do not list explicitly. 

\section{Conclusions}\label{sec_con}
In this paper, we have studied heterotic line bundle models on elliptically fibered CY three-folds. Standard heterotic model building requires a non-trivial first fundamental group of the CY three-fold which is normally realised by starting with a simply-connected CY three-fold $X$ with a freely-acting symmetry $\Gamma$ and then taking the quotient $X/\Gamma$. This has led us to study elliptically fibered CY three-folds  with the simplest type of symmetry, a freely-acting involution $\iota_X$. We have realised this involution by combining a half-shift $\iota_E$ on the elliptic fibers with an involution $\iota_B$ on the base $B$ of the fibration. Such elliptic fibrations necessarily have two sections which are exchanged by $\iota_E$. 

We have systematically developed the tools required for heterotic line bundle model building on such manifolds, including the calculation of line bundle cohomology, by expressing all relevant properties in terms of properties of the base $B$.

The choice of base spaces is restricted by requiring a generically smooth Weierstrass elliptic CY three-fold $X$ and a fixed point free involution $\iota_X$. The latter can be realised if the involution $\iota_B$ on the base has at most fixed points. In this paper, we have focused on toric two-fold base spaces $B$ and, from this class, the above requirements single out six spaces, represented by the reflexive polygons shown in Fig.~\ref{fig:basespaces}.

For those six base spaces, we have systematically searched for physically promising models, that is models based on rank five line bundles sums $V$, with each constituent line bundle being $\iota_X$ invariant, and a total chiral asymmetry of six. After taking the quotient by $\iota_X$ and including a Wilson line these give rise to theories with the standard model group and three chiral families. Over the six possible base manifolds, we have found a total of $442$ models of this kind and we have computed the complete spectrum for all these models. 

A generic feature is the presence of vector-like states, specifically at least one ${\bf 10}$--$\overline{\bf 10}$ pair and at least $20$ ${\bf 5}$--$\bar{\bf 5}$ pairs in each case. In this respect, the results are very different from the ones obtained for line bundle models on complete intersection CY three-folds~\cite{Anderson:2011ns,Anderson:2012yf,Anderson:2013xka}, where vector-like pairs were absent for most models with the correct chiral asymmetry. The underlying reason for this difference seems to be the different structure of line bundle cohomology for the two constructions. Complete intersection CY three-folds are defined in ambient spaces consisting of products of projective spaces. They inherit to some degree the relatively simply structure of line bundle cohomology on projective spaces, where at most one cohomology of a line bundle can be non-vanishing. On the other hand, for the elliptically fibered CY three-folds considered in this paper, line bundle cohomology is determined from line bundle cohomology on the toric two-fold base (or, alternatively, from line bundle cohomology on the ambient toric four-fold), which tends to be more complicated as compared to projective spaces. 

The presence of vector-like states does not mean that our models are phenomenologically ruled out. All models contain standard model singlet fields (which should be interpreted as bundle moduli) and, depending on details, may allow for superpotential couplings of the form ${\bf 1}\,{\bf 10}\,\overline{\bf 10}$ or ${\bf 1}\,{\bf 5}\,\bar{\bf 5}$ (or similar, higher-dimensional operators with multiple singlet insertions). Non-trivial singlet vacuum expectation values which correspond to deformations of the bundle away from a line bundle sum and to a bundle with non-Abelian structure group can then generate masses for the vector-like pairs and remove these states from the low-energy spectrum. A detailed study of this is beyond the scope of the present paper and a subject of future work. Specifically, it would be interesting to check if all unwanted vector-like states can be removed in this way while the desired pair of Higgs doublets can be kept light at the same time. 

We have also seen that chiral line bundle models on elliptically fibered CY three-folds necessarily have a bundle $V$ which restricts non-trivially to the fibers.  Since heterotic F-theory duality is formulated in terms of spectral cover bundles which, by construction, are flat on the fibers it is not clear whether the models found in this paper have an F-theory dual. It would be interesting to investigate the possible relation to F-theory further and work along these lines is in progress.

It is likely that our model building strategy can be generalised to base spaces which are not weak Fano. Besides finding appropriate involutions on those base spaces, this would require us to find an extension of the half-shift involution $\iota_X$ to the reducible fibers present in such models. As such models have more divisors than those associated with sections and divisors of the base, the set of possible line bundles and the Mori cone are larger and more complicated. In practice, the realisation of elliptically fibered CY three-folds as hypersurfaces in toric four-folds, as described in Appendix \ref{app:toric2sectionmodels}, may provide the appropriate framework for such a generalisation to base spaces which are not weak Fano. Pursuing this is an interesting direction for future work.

\section*{Acknowledgements}
We thank Evgeny Buchbinder for collaboration at an early stage of this work, and Fabian Ruehle for helpful discussions. A.~L.~and A.~B.~would like to acknowledge support by the STFC grant~ST/L000474/1.  C.~R.~B. is supported by an STFC studentship. A.~L.~is also partially supported by the EPSRC network grant EP/N007158/1.
\newpage

\appendix

\section{Elliptic fibrations and half-shifts}
\label{ell_curve}
In this appendix we review the required mathematics for elliptically fibered CY three-folds with a freely-acting involution and two sections. Much of the material is not new but can be found in various places in the literature~\cite{Donagi:1999ez,Andreas:1999ty}. 

\subsection{The group law on elliptic curves and half-shifts}
\label{app:grouplaw}

An elliptic curve $E$ can be described by the polynomial equation,
\be\label{eq:weierappendix}
z y^2=x^3+fxz^2+gz^3 \, ,
\ee
where $x,y,z$ are homogeneous coordinates on $\mathbb{P}^2$ and $f$, $g$ are complex constants. This does indeed describe a smooth curve as long as the discriminant, defined by
\begin{equation}
 \Delta=4f^3+27g^2 \label{disE}
\end{equation}
is non-vanishing.

If we denote the nowhere vanishing $(1,0)$-form on $E$ by $\Omega$ and introduce a basis of one-cycles $A$, $B$ with $A\cdot B=1$ on $E$ the periods can be defined by
\begin{equation}
\begin{aligned}
\tau_A = \int_A \Omega \,, \\
\tau_B = \int_B \Omega \,.
\end{aligned}
\end{equation}
For a suitable normalisation of $\Omega$ we have $\tau_B=1$. The other period,  $\tau=\tau_A$, is  called the modular parameter of the torus. We can introduce the lattice $\Lambda\subset\mathbb{C}$ generated by $1$ and $\tau$ and the Jacobian $\mathbb{C}/\Lambda$ of the elliptic curve. Then, the Abel-Jacobi map $E\rightarrow \mathbb{C}/\Lambda$ is defined by
\begin{equation}\label{eq:abeljacobi}
\begin{aligned}
p \mapsto \int_{\gamma} \Omega\, ,
\end{aligned}
\end{equation}
where $\gamma$ is a path linking $p$ with a given point $p_0\in E$ (conveniently taken as $p_0=[x:y:z] = [0:1:0]$). From now on we work in the patch of $\mathbb{P}^2$ where we can set $z=1$. In this patch, the inverse of the Abel-Jacobi map can be expressed in terms of the Weierstrass $\wp$-function as
\begin{equation}
 w\rightarrow (x,y)=(\wp(w),\tfrac{1}{2}\wp'(w))\; , \label{AJinv}
\end{equation} 
where $w\in\mathbb{C}/\Lambda$~\cite{lang1999complex}.

Via this inverse map, the obvious addition of points in $\mathbb{C}/\Lambda$ can be translated into an addition law on the elliptic curve $E$. Specifically, if we consider three points $w_1,w_2,w_3\in \mathbb{C}/\Lambda$ with $w_3=w_1+w_2$, and introduce the corresponding points $(x_i,y_i)=(\wp(w_i),\tfrac{1}{2}\wp'(w_i))$, where $i=1,2,3$, on the elliptic curve, the addition law on $E$ takes the form
\bea
x_3&=&-x_1-x_2+\left(\frac{y_1-y_2}{x_1-x_2}\right)^2 \, , \label{hsx}\\
y_3&=&-\left(\frac{y_1-y_2}{x_1-x_2}\right)^3+\frac{x_1y_1-x_2y_2+2(x_2y_1-x_1y_2)}{x_1-x_2} \label{hsy}\, .
\eea
For fixed $(x_2,y_2)$, this defines a map from $(x_1,y_1)$ to $(x_3,y_3)$ which, provided the discriminant is non-zero, is an automorphism of the elliptic curve.

Since we are interested in constructing involutions of elliptic CY three-folds it is natural to start with involutions on elliptic curves. In the Jacobian description, $\mathbb{C}/\Lambda$, obvious involutions are half-shifts of the form $w\rightarrow w+\omega_i$, with the three possible choices $\omega_1=\frac{1}{2}$, $\omega_2=\frac{\tau}{2}$ or $\omega_3=\frac{1}{2}+\frac{\tau}{2}$ for the shift. Note that these half-shifts are clearly freely-acting and, due to the quotient by the lattice $\Lambda$, they do indeed square to the identity. Via the map~\eqref{AJinv}, the half-shift points $\omega_i\in\mathbb{C}/\Lambda$ can be translated to certain points $(x,y)$ of the elliptic curve $E$. It turns out that $\wp'(\omega_i)=0$ and, hence, that the resulting points are of the form $(x,y)=(e_i,0)\in E$, where $e_i:=\wp(\omega_i)$. Since the $y$-coordinates of these three points vanish the $x$-coordinates $e_i$ must be zeros of the right-hand side of the Weierstrass equation~\eqref{eq:weierappendix}. 

To be specific, we focus on the first half-shift by $\omega_1$ with corresponding half-shift point $(x,y)=(\alpha,0)$, where $\alpha:=e_1$. Since $\alpha$ must be a zero of the right-hand side of Eq.~\eqref{eq:weierappendix}, the Weierstrass equation (for $z=1$) factorises as
\be\label{eq:factweierapp1}
y^2=(x-\alpha)(x^2+\alpha x+\beta) \, ,
\ee
where $\beta$ is another complex number. Comparison with the standard form~\eqref{eq:weierappendix} of the Weierstrass equation shows that $f=\beta-\alpha^2$ and $g=-\alpha\beta$. The discriminant~\eqref{disE} factorises as well and can be written as 
\be\label{eq:discr_factors}
\Delta= \Delta_1\Delta_2^2 \, ,\qquad \Delta_1=4\beta-\alpha^2\;,\qquad \Delta_2=2\alpha^2+\beta\; . 
\ee
In terms of shifted coordinates, defined as $X:= x-\alpha$, $Y:= y$, the above Weierstrass equation takes the form
\be
Y^2=X(X^2+3\alpha X+\Delta_2) \, . \label{Wshifted}
\ee

Setting $x_2=\alpha$, $y_2=0$, $x_1=x$, $y_1=y$, $x_3=x'$ and $y_3=y'$ in Eqs.~\eqref{hsx}, \eqref{hsy} we can translate the half-shift $w\rightarrow w+\omega_1$ on the Jacobian into a map on the elliptic curve which we denote by $\iota_E$. In the affine $(x,y)$ coordinates this map reads explicitly
\bea
x'&=&-x-\alpha+\left(\frac{y}{x-\alpha}\right)^2 \, , \\
y'&=&-\left(\frac{y}{x-\alpha}\right)^3+y\left(\frac{x+2\alpha}{x-\alpha}\right) \, .
\eea
In terms of the shifted $X$, $Y$ coordinates and their primed counterparts, these transformations can be re-written as
\begin{equation}
X'=\frac{1}{X^2}\left(Y^2-X^2\left(X+3\alpha\right)\right)\;,\qquad Y'=-\frac{YX'}{X}\, .
\end{equation}
Finally, using the Weierstrass equation~\eqref{Wshifted} to replace $Y^2$ in the equation for $X'$, the map $\iota_E$ can be cast into the simple form
\be
\iota_E\,:\quad X\rightarrow X'= \frac{\Delta_2}{X} \,, \quad Y\rightarrow Y'=-\frac{\Delta_2Y}{X^2} \,.
\label{eq:halfshiftapp}
\ee
Clearly, since the half-shift $w\rightarrow w+\omega_1$ is freely-acting on the Jacobian, the action of $\iota_E$ is also free, as long as the elliptic curve is smooth. For later purposes, we note that, in the smooth case, $\iota_E$ maps the point $[x:y:z]=[\alpha:0:1]$ into the point $[x:y:z]=[0:1:0]$ or, equivalently,
\begin{equation}
 \iota_E([0:1:0])=[\alpha:0:1]\; . \label{secmap}
\end{equation}
To see this we have to convert the two points into affine coordinates on the patch where $z=1$. The point $[\alpha:0:1]\in\mathbb{P}^2$ then corresponds to $(x,y)=(\alpha,0)$ while the point $[0:1:0]\in\mathbb{P}^2$ is mapped to infinity, $x,y\rightarrow\infty$, with $y^2=x^3$. By taking limits carefully, it can indeed be shown that \eqref{eq:halfshiftapp} exchanges these two affine points.

Let us discuss what happens when the elliptic curve becomes singular, that is, when $\Delta_1=0$ or $\Delta_2=0$. If $\Delta_2=0$, the Weierstrass equation~\eqref{eq:factweierapp1} acquires a double root at $x=\alpha$ (that is, the second factor on the right-hand side of Eq.~\eqref{eq:factweierapp1} develops a root at $\alpha$). In this case, the transformation~\eqref{eq:halfshiftapp} clearly degenerates badly as all points $(X,Y)$ are mapped to the origin $(X',Y')=(0,0)$. As we will explain, in the context of elliptic fibrations, such fibers will be blown-up and the degeneracy of the transformation will be removed in this way.

If $\Delta_1=0$, the Weierstrass equation~\eqref{eq:factweierapp1} develops a double root at $x=-\alpha/2$,  solely from the second factor on the right-hand side. In this case, $\Delta_2 = \frac{9}{4} \alpha^2$, so that the points $(X,Y)=(\tfrac32 \alpha,0)$ and $(X,Y)=(-\tfrac32 \alpha,0)$ are fixed. The second of these points is indeed on the elliptic curve $E$, as can be verified by inserting into Eq.~\eqref{Wshifted}, and, hence, $\iota_E$ is not fixed point free if $\Delta_1=0$.

\subsection{Elliptically fibered CY three-folds with involutions}\label{fibrtwosec_appendix}
We would now like to apply the discussion of the previous sub-section to the construction of elliptically fibered CY three-folds, $X$,  with an involution $\iota_X$. The idea is to construct $\iota_X$ by combining the above half-shift $\iota_E$ on the elliptic fibers with a suitable involution $\iota_B$ on the base $B$ of the fibration.

We start with an elliptically fibered CY three-fold $X$ with two-fold base $B$, projection $\pi:X\rightarrow B$ and section $\sigma:B\rightarrow X$. Each elliptic fiber $E_b=\pi^{-1}(b)$ over $b\in B$ is described by a Weierstrass equation
\be
zy^2=x^3+fxz^2+gz^3 \, , \label{stdw}
\ee
where $x,y,z$ are now sections $x\in\Gamma(K_B^{-2})$, $y\in\Gamma(K_B^{-3})$, $z\in\Gamma({\cal O}_B)$ and $f$, $g$ are sections $f\in \Gamma(K_B^{-4})$, $g \in\Gamma(K_B^{-6})$, with $K_B$ the canonical bundle of the base. Then, the section $\sigma$  is located at the point $[x:y:z]=[0:1:0]$ in each fiber. 

We would now like to consider a situation where a half-shift $\iota_E$ acts on every (smooth) elliptic fiber as in Eq.~\eqref{eq:halfshiftapp}. From the previous sub-section we know that this leads to a factorisation of the Weierstrass equation, namely
\be\label{eq:factglobalweierapp}
zy^2=(x-\alpha z)(x^2+\alpha xz+\beta z^2) \, ,
\ee
where $\alpha$ and $\beta$ should now be seen as sections of $K_B^{-2}$ and $K_B^{-4}$, respectively. This relates to the standard Weierstrass form~\eqref{stdw} via 
\be
f=\beta-\alpha^2\;,\quad g=-\alpha\beta \; .
\ee
As we have seen before, the discriminant factors as
\be\label{eq:discr_factors1}
\Delta= \Delta_1\Delta_2^2 \, ,\qquad \Delta_1=4\beta-\alpha^2\;,\qquad \Delta_2=2\alpha^2+\beta\; . 
\ee
As a consequence, the discriminant locus on the base $B$ consists of the two components $\{b\in B\,|\,\Delta_1(b)=0\}$ and $\{b\in B\,|\,\Delta_2(b)=0\}$.

The half-shift on the fibers leads to a second section, $\zeta:=\iota_E\circ\sigma$ which, in view of Eq.~\eqref{secmap}, is located at $[x:y:z]=[\alpha:0:1]$. We are, therefore, considering elliptically fibered CY three-folds with two sections, which are explicitly given by
\begin{equation}
 \sigma(b)=(b,[0:1:0])\;,\qquad \zeta(b)=(b,[\alpha:0:1])\; .
\end{equation} 
Note that the second section $\zeta$ does indeed take values on the elliptic curve~\eqref{eq:factglobalweierapp}, due to the factorisation of the equation. Hence, the factorisation with holomorphic sections $\alpha$ and $\beta$ is crucial for the existence of a second section.

We should now discuss the singular fibers which arise over the discriminant locus and the fate of the involution $\iota_E$ over these fibers. We begin with the component of the discriminant locus defined by $\Delta_2=0$. As can be seen from Eq.~\eqref{Wshifted}, the three-fold has an $A_1$ singularity at $X=Y=0$ over this locus which needs to be resolved in order to arrive at a smooth CY three-fold. 

A convenient technique to find crepant resolutions of such singularities is to first promote $\Delta_2$ to a coordinate of the ambient space, which then has to satisfy the equation
\be
\Delta_2 = 2\alpha^2+\beta \, . \label{Delta21}
\ee
The blow-up is realised by introducing the new coordinates $\h{X}$,  $\h{Y}$,  $\h{\Delta}_2$ and $\xi$, excising the locus $\{\h{X} = \h{Y} = \h{\Delta}_2\}$ and modding out by a $\C^*$ action with weights
\be
\begin{array}{|c|c|c|c|}
\h{X} & \h{Y} & \h{\Delta}_2 &\xi \\
\hline
1 & 1 & 1 & -1
\end{array}
\ee
These new coordinates relate to the old ones as
\bea
X=\h{X} \xi \, , ~~ Y=\h{Y} \xi \, , ~~ \Delta_2=\h{\Delta}_2 \xi \, .
\eea
After the proper transform, the Weierstrass equation~\eqref{Wshifted} and Eq.~\eqref{Delta21}, turn into
\begin{equation}
\h{Y}^2=\h{X}\left(\h{X}^2\xi+3\alpha \h{X}+\h{\Delta}_2\right)\;,\quad \h{\Delta}_2\xi =2\alpha^2+\beta \, , \label{newCY}
\end{equation}
which describe a smooth CY three-fold. Note that due to the special form of Eq.~\eqref{Wshifted}, the proper transform removes two powers of $\xi$, so that the blow-up is crepant. The previously singular fibers over $\Delta_2 = 0$ have now been replaced by two irreducible components, $\h{\Delta}_2 = 0$ and $\xi = 0$, both of which are $\P^1$s. These two $\P^1$'s are explicitly given by
\bea
\h{Y}^2&=&\h{X}^2\left(\h{X}\xi+3\alpha\right) ~~~~\, \text{for}~\h{\Delta}_2=0 \, , \\
\h{Y}^2&=&\h{X}\left(3\alpha\h{X}+\h{\Delta}_2\right) ~~ \text{for}~\xi=0 \, ,
\eea
and they evidently touch in the two points
\begin{equation}
 \frac{\h{Y}}{\h{X}}=\pm\sqrt{3\alpha}\; . \label{touch}
\end{equation}

What remains to be discussed is the action of the half-shift \eqref{eq:halfshiftapp} on the blown-up space. First note that as long as $\Delta_2\neq0$, we can set $\xi=1$ and our equations return to the ones before the blow-up, so that our earlier discussion applies. This means that the action of $\iota_E$ remains free for fibers which are away from the discriminant locus and that it has fixed points on fibers over $\Delta_1 = 0$. 

In order to discuss the situation for fibers over the locus $\Delta_2 = 0$, we first need to properly define the half-shift in terms of the new coordinates  $\h{X}$,  $\h{Y}$,  $\h{\Delta}_2$ and $\xi$. Of course, we wish to preserve the action of the map on $X=\xi\h{X}$ and $Y=\xi\h{Y}$, which should still be given by Eq.~\eqref{eq:halfshiftapp}. Translating these equations into the new coordinates implies
\begin{equation}
\xi'\hat{X}'   \overset{!}{=} \frac{\Delta_2}{X}=\frac{\h{\Delta}_2}{\h{X}}\; ,\qquad
\xi'\hat{Y}'   \overset{!}{=}  - \frac{\Delta_2 Y}{X^2} = - \frac{\h{\Delta}_2\h{Y}}{\h{X}^2} \; . \label{oldtrafo}
\end{equation}
We also know that, away from the discriminant locus, the half-shift exchanges the two sections and that, for the reducible fibers over the locus $\Delta_2=0$, each $\mathbb{P}^1$ component is met by one of the sections. Since the two $\mathbb{P}^1$'s are given by $\hat{\Delta}_2=0$ and $\xi=0$, respectively, this suggests that $\iota_E$ should exchange the coordinates $\hat{\Delta}_2$ and $\xi$. 
Hence the appropriate generalisation of the half-shift involution to the blown-up space is
\be\label{eq:involution_buspace}
\iota_E:\quad\begin{pmatrix} \h{X} \\ \h{Y} \\ \h{\Delta}_2 \\ \h{\xi} \end{pmatrix}
\longrightarrow
\begin{pmatrix} \frac{1}{\h{X}} \\ -\frac{\h{Y}}{\h{X}^2} \\ \h{\xi} \\  \h{\Delta}_2 \end{pmatrix} \, .
\ee
This transformation respects Eq.~\eqref{oldtrafo} and it exchanges the two $\mathbb{P}^1$'s, defined by $\hat{\Delta}_2=0$ and $\xi=0$, in the reducible fibers.

What about the fixed points for this refined version of the involution $\iota_E$? By construction, Eq.~\eqref{eq:involution_buspace} reduces to the previous action for all fibers away from the locus $\Delta_2=0$. Hence, it is fixed point free on fibers away from the discriminant locus and it has fixed points on fibers over the locus $\Delta_1=0$ (certainly as long as $\Delta_2\neq 0$). For fibers over the locus $\Delta_2 = 0$, the map~\eqref{eq:involution_buspace} exchanges the two fiber components $\xi=0$ and $\h{\Delta}_2=0$. These two $\mathbb{P}^1$ components touch in the two points~\eqref{touch} which are also swapped by the map \eqref{eq:involution_buspace}. Hence, as long as these two points are different the action of $\iota_E$ on fibers over the locus $\Delta_2=0$ is fixed point free. The two points coincide precisely when $\Delta_2 = \Delta_1= 0$ and the resulting fixed point is what we expect on fibers over the locus $\Delta_1=0$. In summary, the involution $\iota_E$, as defined by Eq.~\eqref{eq:involution_buspace}, is fixed point free over all fibers with $\Delta_1\neq 0$ and it has fix points over fibers with $\Delta_1=0$. 

It remains to discuss the two sections $\sigma$ and $\zeta$ for the blown-up version of the CY three-fold. The zero section $\sigma$ is specified by mapping a point in the base to the point
\be
\h{X}\to\infty\;,\quad \h{Y}\to\infty\;,\quad \h{\Delta}_2=2\alpha^2+\beta\;,\quad \h{\xi}=1 \, , \label{ps}
\ee
where $\h{X}$ and $\h{Y}$ are taken to infinity such that $\h{Y}^2=\h{X}^3\h{\xi}$ is satisfied along the path. (This is required for the first equation in~\eqref{newCY} to be satisfied in the limit.) The second section, $\zeta$, maps a point in the base to
\be
\h{X}\to0\,,~\h{Y}\to0\,,~\h{\Delta}_2=1\,,~\h{\xi}=2\alpha^2+\beta \, , \label{pz}
\ee
where $\h{X}$ and $\h{Y}$ are taken to zero such that $\h{Y}^2=\h{X}\h{\Delta}_2$ is satisfied along the path. (This is required for the first equation in~\eqref{newCY} to be satisfied in the limit.) It is straightforward to check that the involution~\eqref{eq:involution_buspace} swaps the two section points~\eqref{ps} and \eqref{pz} and, hence, swaps the two sections $\sigma$ and $\zeta$.

\subsection{Alternative realisation of elliptic Calabi-Yau three-folds with half-shifts}\label{app:toric2sectionmodels}

There exists another presentation of elliptic Calabi-Yau three-folds with a second section at the two-torsion point which uses a different embedding of the elliptic fiber. Embedding the elliptic fiber in the Hirzebruch surface $\mathbb{F}_2$ (which can also be found as a blow-up of $\P^2_{112}$), an elliptic fibration with two sections has the realisation
\begin{equation}\label{eq:ellfib2secttoric}
x_1^2 = x_2^2 (x_3^4 + b_2 x_3^2 x_4^2 + b_3 x_3 x_4^3 + b_4 x_4^4)  \, .
\end{equation}
Here, the $x_i$ are the homogeneous coordinates of a toric variety with weight system 
\begin{equation}
 \begin{array}{cccc}
  x_1 &x_2& x_3& x_4 \\
  \hline
  1 & 1 & 0 & 0 \\
  2 & 0 & 1 & 1
 \end{array}
\end{equation}
and Stanley-Reisner ideal generated by $\{x_1,x_2\},\{x_3,x_4\}$. Note that we may always find this form by appropriately redefining the coordinates $x_1$ and $x_3$ of a general hypersurface in $\mathbb{F}_2$. In order for a fibration with the above elliptic curve to form a CY space, 
the $b_i$ must be sections of the line bundles $-K_B^{\otimes i}$. 

The two sections of such a model are located at $x_4 = x_1 \pm x_2 x_3^2 = 0$. Choosing one of the two as the zero section, the other section is located at the two-torsion point if we set $b_3 \equiv 0$. The involution corresponding to the half-shift is then simply given by 
\begin{equation}
(x_1,x_2,x_3,x_4)\rightarrow (-x_1,x_2,x_3,-x_4)
\end{equation}
and the discriminant becomes the same as \eqref{eq:discr_factors1}, when using the replacements $b_2 = -6i\alpha$ and $b_4 = \Delta_1 = 4\beta - \alpha^2$. This action is free as long as $b_4 = \Delta_1 \neq 0$.

In contrast to the realisation via a specialised Weierstrass model, no further blow-ups are required in order to find a smooth Calabi-Yau three-fold $X$. Furthermore, for a toric base space $B$, this presentation allows a description of $X$ as a toric hypersurface as follows. Let us denote the rays of the fan of the toric base $B$ by ${\bf w}_i \in \mathbb{Z}^2$. If the convex hull of the vectors
\begin{equation}
\left(\begin{array}{r}
1 \\ 0 \\ {\bf 0 } 
\end{array}\right)\, ,
\left(\begin{array}{r}
-1 \\ 0 \\ {\bf 0 } 
\end{array}\right)\, ,\left(\begin{array}{r}
0 \\ -1 \\ {\bf 0 } 
\end{array}\right)\, ,
\left(\begin{array}{r}
-2 \\ -1 \\ {\bf 0 } 
\end{array}\right)\, ,
\left(\begin{array}{r}
-2 \\ -1 \\ {\bf w }_i 
\end{array}\right)\, ,
\end{equation}
forms a reflexive polytope $\Delta^\circ$, we can realise \eqref{eq:ellfib2secttoric} as a generic toric hypersurface which sits in a fibration of $\mathbb{F}_2$ over $B$. For the base spaces we consider, which are toric weak Fano two-folds, this condition is always met. In order to find candidate free involutions, we may then restrict to cases where $b_3 = 0$. In this form, our examples should also be found using the algorithm of Ref.~\cite{Braun:2017juz}.

A further upshot of this alternative realisation is that it allows for an easy alternative method to compute line bundle cohomologies. Instead of exploiting the elliptic fibration, as described in Appendix~\ref{app:cohomtwosections}, we may simply compute cohomologies on the toric ambient four-fold, using the code \cohomCalg \cite{CohomOfLineBundles:Algorithm,cohomCalg:Implementation}, and then apply the Koszul sequence to find the cohomologies on $X$. We have used this alternative method to check our results for line bundle cohomology.

\subsection{Intersection theory on CY three-folds with involutions}\label{app:intersectionson3fold}
The main purpose of this section is to introduce a suitable integral basis of curve and divisor classes on the CY three-fold $X$ in terms of a basis of curve classes on the base $B$ and to study the resulting intersection properties.

We begin by introducing the required objects on the base $B$. These consist of a basis, $\{{\cal C}^i\}$, where $i,j,\ldots=1,\ldots ,h^{1,1}(B)$, of curve classes, its dual basis $\{{\cal C}_i\}$ such that
\begin{equation}
 \mc{C}^i\cdot \mc{C}_j=\delta^i_j\; ,
\end{equation} 
and the various intersection numbers
\begin{equation}
 g_{ij}={\cal C}_i\cdot{\cal C}_j\;,\quad g^{ij}={\cal C}^i\cdot{\cal C}^j\;, \quad \lambda_i=K_B\cdot{\cal C}_i\;,\quad \lambda=\lambda^i\lambda_i\; ,
\end{equation} 
where $K_B$ is the canonical bundle of $B$. Note that $g^{ij}$ is the inverse of $g_{ij}$ and that these two metrics can be used to raise and lower indices. In particular, we have $\lambda^i=g^{ij}\lambda_j$.

Next, we should construct a basis of curve classes on the CY three-fold $X$ which can be done by ``lifting" the above curve classes ${\cal C}^i$ on the base, using the sections $\sigma$ or $\zeta$. There are two further curve classes on $X$ which cannot be obtained in this way, namely the class of the generic fiber and the new class introduced by the blow-up. We denote the class of the generic fiber by $F$ and the new class, represented by the $\xi=0$ component of the reducible fibers, by $N$. The other component, defined by $\hat{\Delta}_2=0$, of the reducible fibers then has the class $F-N$. 
It turns out that, for a general curve class, ${\cal C}$, on the base we have the relation~\cite{Donagi:1999ez}
\be
\zeta({\cal C})=\sigma({\cal C})+({\cal C}\cdot c_1(B))[F-2N]\; . \label{zetasigma}
\ee
This means that the lifts of a base curve by the two sections are linearly related and that we can focus on one of the sections for the purpose of constructing a basis of curve classes on $X$. With this in mind we introduce a basis $\{C^I\}$, where $I=(0,\hat{0},i)$ and $i=1,\ldots ,h^{1,1}(B)$, of curve classes on $X$ by setting
\begin{equation}
 C^0=F-N\;,\quad C^{\hat{0}}=N\;,\quad C^i=\sigma({\cal C}^i)-\lambda^i(F-N)\; .
\end{equation} 
(The last term in the definition of $C^i$ has been included for later convenience.) Then, an arbitrary second homology class in $H_2(X,\mbb{Z})$ can then be written as $C=n_IC^I$, where $n_I\in\mbb{Z}$.

To obtain an integral basis of divisor classes on $X$, we proceed as follows. We can find divisors of $X$ by lifting curves of the base using the inverse projection map, $\pi^{-1}$, of the fibration. There are only two further divisor classes, the images $\sigma(B)$ and $\zeta(B)$ of the base, which cannot be obtained in this way. Hence, we introduce the basis $\{D_I\}$, where $I=(0,\hat{0},i)$ and $i=1,\ldots ,h^{1,1}(B)$, by
\begin{equation}
 D_0=\sigma(B)\;,\quad D_{\hat{0}}=\zeta(B)\;,\quad D_i=\pi^{-1}({\cal C}_i)\; .
\end{equation} 
As Table~\ref{tab_CDisec} shows, this basis is dual to the above basis of curve classes, that is
\begin{equation}
 D_I\cdot C^J=\delta_I^J\; .
\end{equation} 
\begin{table}[h]
\begin{center}
\begin{tabular}{|C||C|C|C|}\hline
				& D_{0}=\sigma(B) & D_{\hat{0}}=\zeta(B)& \pi^{-1}({\cal C}')	\\ \hline\hline
C^{0}=F-N        	& 1					& 0				& 0				\\\hline
C^{\hat{0}}=N 			& 0					& 1				& 0				\\\hline
\sigma({\cal C}) 	& {\cal C}\cdot K_B		& 0                    	& {\cal C}\cdot{\cal C}'	\\\hline
\zeta({\cal C}) 		& 0                     		& {\cal C}\cdot K_B	& {\cal C}\cdot{\cal C}'	\\\hline
\end{tabular}
\caption{\it Intersection properties of curve classes and divisor classes. Here, ${\cal C}$ and ${\cal C}'$ are two curve classes on the base.}\label{tab_CDisec}
\end{center}
\end{table}
We also require the intersections of two divisors which are summarised in Table~\ref{tab_DDisec}.
\begin{table}[h]
\begin{center}
\begin{tabular}{|C||C|C|C|}\hline
				& D_{0}=\sigma(B)       	& D_{\hat{0}}=\zeta(B)		& \pi^{-1}({\cal C}')	\\ \hline\hline
D_{0}=\sigma(B)		& \sigma(K_B)			& 0				& \sigma({\cal C}')	\\\hline
D_{\hat{0}}=\zeta(B)    	& 0					& \zeta(K_B)		&\zeta({\cal C}')	\\\hline
\pi^{-1}({\cal C}) 	& \sigma({\cal C}) 		& \zeta({\cal C}) 	& ({\cal C}\cdot{\cal C}')F	\\\hline
\end{tabular}
\caption{\it Intersection properties of two divisor classes. Here, ${\cal C}$ and ${\cal C}'$ are two curve classes on the base.}\label{tab_DDisec}
\end{center}
\end{table}
Combining the information from Table~\ref{tab_CDisec} and Table~\ref{tab_DDisec}, we can work out the triple intersection numbers 
\begin{equation}
 d_{IJK}:=D_I\cdot D_J\cdot D_K\; ,
\end{equation}
which are explicitly given by 
\begin{equation}
d_{000}=d_{\hat{0}\hat{0}\hat{0}}=\lambda\;, \quad d_{00i}=d_{\hat{0}\hat{0}i}=\lambda_i \;, \quad d_{0ij}=d_{\hat{0}ij}=g_{ij}\;,
\end{equation}
with all other components fixed by symmetry or else vanishing. 

Finally, it can be shown~\cite{Donagi:1999ez,Andreas:1999ty} that on an elliptically fibered Calabi-Yau $X$ with two sections, the second Chern class $c_2(X)$ and the Euler number $\chi(X)$ can be expressed in terms of properties of the base as
\bea
 c_2(X)&=&12\sigma(c_1(B))+(c_2(B)+11c_1(B)^2)(F-N)+(c_2(B)-c_1(B)^2)N  \\
&=& (c_2(B)-\lambda)(C^0+C^{\hat{0}})-12\lambda_iC^i\;,\\
\chi(X)&=&-36\int_Bc_1(B)^2 =-36\lambda \, .
\eea

\subsection{Mori cone and K\"ahler cone}
\label{mori_kahl_cones_appendix}
For the purpose of checking the heterotic anomaly condition we need to know the Mori cone of the CY three-fold, after the blow-up. As we will see, the Mori cone ${\cal M}_X$ of $X$ can be expressed in terms of the Mori cone ${\cal M}_B$ of the base. 

First we note that for two effective curve classes ${\cal C},{\cal C}'\in{\cal M}_B$ in the base, the curve class
\be
C=m(F-N)+nN+\,\sigma(\mc{C})+\,\zeta(\mc{C'})\;, \label{MC}
\ee
where $n,m\in\mathbb{Z}^{\geq 0}$, must be in the Mori cone of $X$. We would now like to argue that all effective curve classes on $X$ are, in fact, obtained in this way. 

It is sufficient to show that all irreducible effective classes can be represented in the form~\eqref{MC} (since an extremal ray of the cone of effective classes is an irreducible effective class). We write a general class $\hat{C}\in H_2(X,\mbb{Z})$ as,
\be
 \hat{C}=\sigma(\mc{C})+\hat{m} (F-N)+\hat{n} N \, ,
\ee
with $\mc{C}$ an element of $H_2(B,\mbb{Z})$ and $\hat{m}, \hat{n} \in \mbb{Z}$. (Note that, by virtue of the relation~\eqref{zetasigma}, we do not need to include a term $\zeta(\mc{C}')$ in this expression.) Let us now require that $\hat{C}$ is effective and irreducible. Effectiveness means that the projection of $\hat{C}$ onto the base, which is simply ${\cal C}$, is an effective class on the base. Irreducibility of $\hat{C}$ implies, in particular, that a curve in this class intersects non-negatively any effective divisor in which it is not contained, or more precisely whenever the generic element of the curve class does not sit inside the generic element of the divisor class. If $\hat{C}$ is not contained in $\sigma(B)$ or $\zeta(B)$, then exploiting this fact leads to the conditions
\begin{equation}
 0 \leq \sigma(B) \cdot \hat{C} = \hat{m} + K_B \cdot \mc{C}\; ,\qquad 0 \leq \zeta(B) \cdot C = \hat{n}\; .
\end{equation} 
We have seen that the curve ${\cal C}$ is effective and, for the base spaces we consider, we have $-K_B \cdot \,\mc{C} \geq 0 $ for all effective curves. Hence, we conclude that $\hat{m} \geq 0$ and $\hat{n} \geq 0$. In conclusion, if the effective (and irreducible) class $\hat{C}$ is not contained in $\sigma(B)$ or $\zeta(B)$ then it is of the form~\eqref{MC}. Now suppose $\hat{C
}$ is contained in $\sigma(B)$. Then it follows that $\hat{C} = \sigma(\mc{C})$ and, since ${\cal C}$ is effective on the base, this is already of the form~\eqref{MC}. A similar argument applies if $\hat{C}$ is contained in $\zeta(B)$. In this case, $\hat{C}=\zeta(\mc{C'})$ with an effective ${\cal C}'$, which is again of the form~\eqref{MC}. In summary, the Mori cone of the elliptically fibered CY three fold $X$ consists of classes of the form~\eqref{MC} and can, hence, be written as
\begin{equation}
 {\cal M}_X=\left\{m(F-N)+nN+\sigma({\cal C})+\zeta({\cal C}')\,|\,m,n\in\mathbb{Z}^{\geq 0}\;,\;\; {\cal C},{\cal C}'\in{\cal M}_B\right\}\; , \label{MoriX1}
\end{equation}
where ${\cal M}_B$ is the Mori cone of the base $B$. 

In the context of line bundle model building, we will need to check if the slope of line bundles vanishes somewhere in the K\"ahler cone, ${\cal K}_X$, of the CY three-fold $X$ and, hence, we need an explicit description of ${\cal K}_X$. For the manifolds in question the K\"ahler cone is the dual of the Mori cone (see for example Theorem 1.4.9 in \cite{lazarsfeld2004positivity}). Writing a general K\"ahler form as $J=t^ID_I$, where ${\bf t}=(t^I)$ are the K\"ahler moduli relative to a basis $\{D_I\}$ of divisor classes, and writing a general curve class as $n_IC^I$ relative to a dual basis $\{C^I\}$ (which satisfies $C_I\cdot D^J=\delta_I^J$), the K\"ahler cone can be expressed as
\begin{equation}
 {\cal K}_X\cong\{{\bf t}\,|\,{\bf t}\cdot{\bf n}\geq 0\mbox{ for all }n_IC^I\in{\cal M}_X\}\; ,
\end{equation} 
and can, hence, be obtained from the Mori cone of $X$.

\section{Base spaces}
\label{bases_and_involutions}
So far, we have expressed all relevant properties of elliptically fibered CY three-folds with two sections in terms of corresponding properties of the base. In this appendix, we will focus on the six suitable toric base spaces which we have identified. More specifically, from the $61,359$ possible toric two-folds classified in Ref.~\cite{Morrison:2012js} only the $16$ cases identified in Refs.~\cite{Morrison:2012js,Halverson:2015jua} lead to generically smooth elliptic fibrations in Weierstrass form. From those $16$ two-folds we have found involutions with at most fixed points (and which can be realised linearly on the homogeneous coordinates) on precisely the aforementioned six spaces. These spaces are denoted by  $B=F_2, F_4, F_7, F_9, F_{13}, F_{15}$ and their associated polygons are shown in Fig.~\ref{fig:basespaces}.

Our main goal is to list all required properties for these six base spaces $B$. This includes simple topological properties such as the canonical bundle $K_B=-c_1(B)$, the second Chern class $c_2(B)$ and the Hodge number $h^{1,1}(B)$. Further, we provide an integral basis $\{{\cal C}^i\}$ of curve classes, where $i,j,\ldots =1,\ldots ,h^{1,1}(B)$, as well as a corresponding dual basis $\{{\cal C}_i\}$ which satisfies
\begin{equation}
 {\cal C}^i\cdot {\cal C}_j=\delta^i_j\; .
\end{equation} 
We also explain how this basis relates to the divisors $[x_i]$, associated to the curves defined by $x_i=0$, where $x_i$ are the coordinates assigned to the edges of the polygons in Fig.~\ref{tab:basespaces}. For this basis, we provide the intersection numbers
\begin{equation}
 g_{ij}:={\cal C}_i\cdot{\cal C}_j\;,\quad \lambda_i:=K_B\cdot{\cal C}_i\;,\quad \lambda:=K_B^2=c_1(B)^2=\lambda^i\lambda_i\; ,
\end{equation} 
and we note that the intersection matrix $g^{ij}:={\cal C}^i\cdot{\cal C}^j$ is the inverse of $g_{ij}$. The metric $g_{ij}$ and its inverse $g^{ij}$ can be used to lower and raise indices so that, for example, $\lambda^i=g^{ij}\lambda_j$. We will also provide a set of generators of the Mori cone ${\cal M}_B$. 

We note that for any of these bases $B$ these properties can be read off from the corresponding polytope in Fig.~\ref{fig:basespaces}. First note that for any vector $\vec{v}$ in the plane of the polytope there is a linear relation between divisors,
\be
\sum_{i=1}^n \left(\vec{x_i}^T \vec{v}\right)[x_i] = 0 \,,
\ee
where $n$ is the number of external points on the polygon, and where $\vec{x_i}$ is the vector of the point on the edge of the polytope. Note that there are two independent such relations, and hence $h^{1,1}(B)=n-2$. To compute intersection numbers, first note that two $[x_i]$ corresponding to adjacent points have mutual intersection number 1. Self-intersections $[x_i]\cdot[x_i]$ can then be computed by combining this fact with the linear relations between divisors. Additionally we can read off the Chern classes,
\be
c_1(B)=-K_B = \sum_{i=1}^n [x_i] \;, \quad c_2(B) = n \;.
\ee
Finally, the Mori cone is generated by the collection of the $[x_i]$.

We have searched for involutions which can be linearly realised on the homegeneous coordinates and which have at most fixed points. We will explicitly specify the action of these involution $\iota_B$ on the homogeneous coordinates. In addition, we also provide the matrix $I_B$ which describes the action of $\iota_B$ on the curve classes in line with the relation
\begin{equation}
 \iota_B({\cal C}_j)=I^i_{Bj}{\cal C}_i\; .
\end{equation} 
Finally, we indicate the most general form of curve classes ${\cal C}=k^i{\cal C}_i$ which are invariant under $\iota_B$ as a list of constraints on the integers $k^i$ and the resulting dimension, $h^{1,1}_{\rm inv}(B)$, of the sub-space of invariant classes. 

As for the resulting CY three-folds over these base spaces $B$, recall from Eqs.~\eqref{c2X}, \eqref{chiX} and \eqref{MoriX} that we can express the second Chern class $c_2(X)$, the Euler characteristic $\chi(X)$  and the Mori cone $\mc{M}_X$ in terms of properties of the base $B$. We also recall that the {\kahl} cone of the three-fold $X$ is obtained as the dual cone to the Mori cone. Concretely this means one can obtain the inequalities constraining  the {\kahl} moduli by dotting the vector $(t^I)$ of {\kahl} moduli into each of the Mori cone generators (seeing the latter as vectors relative to the basis $\mc{C}^I$). Finally, by noting that $h^{1,1}(X)=h^{1,1}(B)+2$ we can also obtain the Hodge numbers of the three-fold. Using these results,  we will list explicitly, for each base space $B$, the properties of the associated elliptically fibered CY three-fold $X$ with two sections over that base.

\subsection{Base $B=F_2=\mbb{P}^1 \times \mbb{P}^1$}
Our first and simplest base space is $B=F_2= \mbb{P}^1 \times \mbb{P}^1$ with the basic topological characteristics
\begin{equation}
 K_B=-c_1(B)=-2(l_1+l_2)\;,\quad c_2(B)=4\;,\quad h^{1,1}(B)=2\; .
\end{equation} 
Here, $l_1$ and $l_2$ are the divisor classes which correspond to one of the $\mathbb{P}^1$ factors times a point in the other. (These are the pullbacks of the two $\mathbb{P}^1$ hyperplane classes to the product space.) In terms of the coordinates indicated in Fig.~\ref{fig:basespaces}, they can also be written as $l_1=[x_1]$ and $l_2=[x_2]$. 

The intersections of $l_1$ and $l_2$ are clearly given by $l_1\cdot l_1=l_2\cdot l_2=0$ and $l_1\cdot l_2=1$ and, hence, we can introduce a basis of curve classes and its dual by
\be
({\cal C}^i)=(l_1,l_2)\; ,\quad ({\cal C}_i)=(l_2,l_1) \; .
\ee 
The relevant intersection numbers are then given by
\begin{equation}
 (g_{ij})=\left(\begin{array}{ll}0&1\\1&0\end{array}\right)\;,\quad (\lambda_i)=(-2,-2)\;,\quad \lambda=8\; .
\end{equation} 
The Mori cone can be written as
\begin{equation}
 {\cal M}_B=\langle l_1,l_2\rangle\; ,
\end{equation}
where the angle brackets are used to denote the list of generators. With the homogenous coordinates $(x_1,x_2,x_3,x_4)$ as in Fig.~\ref{fig:basespaces} (where $(x_1,x_3)$ correspond to one $\mathbb{P}^1$ factor and $(x_2,x_4)$ to the other), the action of the involution takes the form
\begin{equation}
 \iota_B(x_1,x_2,x_3,x_4)=(x_1,x_2,-x_3,-x_4)\; .
\end{equation}
Both curve classes $l_1$ and $l_2$ are invariant under $\iota_B$ so that $I_B=\mathbbm{1}_2$. Consequently, every curve class on $B$ is $\iota_B$ invariant and we have $h^{1,1}_{\rm inv}(B)=2$.

The resulting properties of the CY three-fold $X$ over this base are
\be
h^{1,1}(X)=4 \;, \quad h^{2,1}(X)=148 \;, \quad \chi(X)=-288 \;, \quad c_2(X)=-4(C^0+C^{\hat{0}})+24(C^1+C^2) \; ,
\ee
\be
\mc{M}_X = \langle {C^{\h{0}},C^0,-2C^0+C^1,-2C^0+C^2,C^1-2C^{\h{0}},C^2-2C^{\h{0}}} \rangle \; .
\ee

\subsection{Base $B=F_4=\mathbb{F}_2$}
This base has the basic properties
\begin{equation}
 K_B=-c_1(B)=-2[x_2]\;,\quad c_2(B)=4\;,\quad h^{1,1}(B)=2\; .
\end{equation} 
In terms of the coordinates in Fig.~\ref{fig:basespaces}, we can introduce a basis of curve classes by ${\cal C}^1=[x_2]$ and ${\cal C}^2=[x_3]$ with dual basis ${\cal C}_1=[x_3]$ and ${\cal C}_2=[x_2-2x_3]$. The corresponding intersection numbers are
\begin{equation}
 (g_{ij})=\left(\begin{array}{rr}0&1\\1&-2\end{array}\right)\;,\quad 
 (\lambda_i)=(-2,0)\;,\quad \lambda=8\; ,
\end{equation} 
and the Mori cone is given by
\begin{equation}
 {\cal M}_B=\langle [x_2],[x_3],[x_2-2x_3]\rangle\; .
\end{equation} 
For the action of the  involution $\iota_B$ on the homogeneous coordinates we have
\begin{equation}
 \iota_B(x_1,x_2,x_3,x_4)= (x_1,x_2,-x_3,-x_4)\; .
\end{equation}
The resulting transformation on the curve classes is trivial so that $I_B=\mathbbm{1}_2$. Consequently, every curve class on $B$ is $\iota_B$ invariant and, hence, $h^{1,1}_{\rm inv}(B)=2$. 

For the associated CY three-fold this leads to
\be
h^{1,1}(X)=4 \;, \quad h^{2,1}(X)=148 \;, \quad \chi(X)=-288 \;, \quad c_2(X)=-4(C^0+C^{\hat{0}})+24C^1 \; ,
\ee
\be
\mc{M}_X = \langle C^{\h{0}},C^0,C^1,-2C^0+C^2,C^2-2C^{\h{0}} \rangle \; .
\ee

\subsection{Base $B=F_7=dP_3$}
\label{using_delpezzo_appendix}
This space equals the del Pezzo surface $dP_3$ which can be seen as $\mathbb{P}^2$ blown-up in three distinct points. Its basic topological properties  are 
\begin{equation}
 K_B=-c_1(B)=-3l+E_1+E_2+E_3\;,\quad c_2(B)=6\;,\quad h^{1,1}(B)=4\; .
\end{equation}
Here, $l$ is the hyperplace class and $E_a$, where $a=1,2,3$, are the classes of the three blow-ups. In terms of the coordinates in Fig.~\ref{fig:basespaces}, these classes can also be expressed as
\begin{equation}
 l=[x_4]+[x_5]+[x_6]\; ,\quad E_1=[x_6]\;,\quad E_2=[x_2]\;,\quad E_3=[x_4]\; .
\end{equation} 
The intersections are given by the well-known formulae
\be
 l\cdot l=1\,,\quad l\cdot E_a=0\,,\quad E_a\cdot E_b=-\delta_{ab}\; ,
\ee 
so that a suitable choice for the basis of curve classes and its dual is given by
\be
({\cal C}^i)=(l,E_a)\, ,\quad ({\cal C}_i)=(l,-E_a) \; .
\ee 
In terms of this basis choice the intersection numbers read
\be
(g_{ij})=\left(\begin{array}{cc}1&0\\0&-{\bf 1}_3\end{array}\right)\,,\quad (\lambda_i)=(-3,1,1,1)\,,\quad \lambda=6\; .
\ee 
The Mori cone is generated by
\begin{equation}
 {\cal M}_B=\langle E_1,E_2,E_3,l-E_1-E_2,l-E_1-E_3,l-E_2-E_3\rangle\; .
\end{equation} 
The action of $\iota_B$ on the homogeneous coordinates
\begin{equation}
\iota_B(x_1,x_2,x_3,x_4,x_5,x_6)=(x_4,x_5,x_6,x_1,x_2,x_3)\; ,
\end{equation}
leads to the action on the curve classes specified by
\be
I_B =
\left(\begin{array}{*{4}{@{}W{\mycolwd}@{}}}
2 & -1 & -1 & -1 \\
1 & 0 & -1 & -1 \\
1 & -1 & 0 & -1 \\
1 & -1 & -1 & 0
\end{array}\right) \; .
\ee
The most general $\iota_B$ invariant curve class $k^i{\cal C}_i$ is then characterised by the constraint
\begin{equation} 
 k^1=k^2+k^3+k^4\; ,
\end{equation} 
so that $h^{1,1}_{\rm inv}(B)=3$.

The associated CY three-fold has the properites
\be
h^{1,1}(X)=6 \;, \quad h^{2,1}(X)=114 \;, \quad \chi(X)=-216 \;, \quad c_2(X)=12(3C^1-C^2-C^3-C^4) \; ,
\ee
\bea
\mc{M}_X &=& \langle C^{\h{0}},C^0,-C^0+C^2,-C^0+C^3,-C^0+C^4,-C^0+C^1-C^2-C^3, \nonumber\\
&& -C^0+C^1-C^2-C^4, -C^0+C^1-C^3-C^4,C^2-C^{\h{0}},C^3-C^{\h{0}},C^4-C^{\h{0}}, \nonumber\\
&& C^1-C^2-C^3-C^{\h{0}},C^1-C^2-C^4-C^{\h{0}},C^1-C^3-C^4-C^{\h{0}} \rangle \;. \label{MCF7}
\eea

\subsection{Base $B=F_9$}
The basic properties of this space are given by
\begin{equation}
 K_B=-c_1(B)=-[x_2]-2[x_3]-[x_4]\;,\quad c_2(B)=6\;,\quad h^{1,1}(B)=4\; ,
\end{equation} 
where $x_1,\ldots ,x_6$ are the homogeneous coordinates as indicated in Fig.~\ref{fig:basespaces}. The standard basis of curve classes and its dual can be introduced as 
\begin{align}
  &{\cal C}^1=[x_2]\;,\quad {\cal C}^2=[x_3]\;,\quad {\cal C}^3=[x_4]\;,\quad {\cal C}^4=[x_5]\\
  &{\cal C}_1=[x_4+x_5-x_2]\;,\quad {\cal C}_2=[x_4+x_5]\\
  &{\cal C}_3=[x_2+x_3-x_4-x_5]\;,\quad {\cal C}_4=[x_2+x_3-x_4-2x_5]\; .
\end{align}  
This leads to the intersections
\be
(g_{ij})=
\left(\begin{array}{*{4}{@{}W{\mycolwd}@{}}}
-1 & 0 & 1 & 1 \\
0 & 0 & 1 & 1 \\
1 & 1 & -1 & -1 \\
1 & 1 & -1 & -2 
\end{array}\right)\;,\quad
(\lambda_i)=(-1,-2,-1,0)\;,\quad
\lambda=6\; .
\ee
The Mori cone is generated by
\begin{equation}
 {\cal M}_B=\langle [x_2],[x_3],[x_4],[x_5],[x_4+x_5-x_2],[x_2+x_3-x_4-2x_5]\rangle\; .
\end{equation}
With the involution $\iota_B$ given by
\begin{equation}
\iota_B(x_1,x_2,x_3,x_4,x_5,x_6)=(x_5,x_4,x_3,x_2,x_1,-x_6)
\end{equation}
the action on the curve classes is specified by
\begin{equation}
I_B=
\left(\begin{array}{*{4}{@{}W{\mycolwd}@{}}}
0 & 0 & 1 & 0 \\
0 & 1 & 0 & 0 \\
1 & 0 & 0 & 0 \\
-1 & 0 & 1 & 1
\end{array}\right) \;,
\end{equation}
leading to $\iota_B$ invariant classes $k^i{\cal C}_i$ characterised by
\begin{equation}
 k^1=k^3\; .
\end{equation} 
This implies $h^{1,1}_{\rm inv}(B)=3$.

For the associated CY three fold we find
\be
h^{1,1}(X)=6 \;, \quad h^{2,1}(X)=114 \;, \quad \chi(X)=-216 \;, \quad c_2(X)=12(C^1+2C^2+C^3) \; ,
\ee
\bea
\mc{M}_X &=& \langle C^{\h{0}},C^0,-C^0-C^1+C^3+C^4,-C^0+C^1,-C^0+C^3,-C^0+C^4, \nonumber\\
&& C^1+C^2-C^3-2C^4,-C^1+C^3+C^4-C^{\h{0}},C^1-C^{\h{0}},C^3-C^{\h{0}},C^4-C^{\h{0}} \rangle \; .
\eea

\subsection{Base $B=F_{13}$}
The basic topological characteristics of this space are
\begin{equation}
 K_B=-c_1(B)=-[x_3]-2[x_4]-[x_5]\;,\quad c_2(B)=8\;,\quad h^{1,1}(B)=6\; .
\end{equation}
and, in terms of the coordinates in Fig.~\ref{fig:basespaces}, a suitable basis of curves is defined by
\begin{equation}
 {\cal C}^i=[x_{i+1}]\;,\quad i=1,\ldots ,6\; .
\end{equation} 
From the intersection numbers
\be
(g_{ij})=
\left(\begin{array}{*{6}{@{}W{\mycolwd}@{}}}
-2 & -1 & 0 & 1 & 2 & 1 \\
-1 & -1 & 0 & 1 & 2 & 1 \\
0 & 0 & 0 & 1 & 2 & 1 \\
1 & 1 & 1 & -1 & -2 & -1 \\
2 & 2 & 2 & -2 & -6 & -3 \\
1 & 1 & 1 & -1 & -3 & -2 
\end{array}\right)\; ,\quad
(\lambda_i)=(0,-1,-2,-1,0,0)\;,\quad 
\lambda=4\; ,
\ee
we can infer the form of the dual basis in terms of classes $[x_i]$ by using the relation ${\cal C}_i=g_{ij}{\cal C}^j$. The Mori cone of this space is generated by
\begin{equation}
 {\cal M}_B=\langle [x_2],\ldots ,[x_7],[x_2+x_3+x_4-x_5-3x_6-2x_7],[-2x_2-x_3+x_5+2x_6+x_7]\rangle\; .
\end{equation}
The involution acts on the homogeneous coordinates as
\begin{equation}
 \iota_B(x_1,x_2,x_3,x_4,x_5,x_6,x_7,x_8)=(x_7,x_6,x_5,x_4,x_3,x_2,x_1,-x_8)\; ,
\end{equation}
leading to an action on the curves classes specified by
 \be
I_B=
\left(\begin{array}{*{6}{@{}W{\mycolwd}@{}}}
0 & 0 & 0 & 0 & 1 & 0 \\
0 & 0 & 0 & 1 & 0 & 0 \\
0 & 0 & 1 & 0 & 0 & 0 \\
0 & 1 & 0 & 0 & 0 & 0 \\
1 & 0 & 0 & 0 & 0 & 0 \\
-2 & -1 & 0 & 1 & 2 & 1
\end{array}\right)\; .
\ee
This means that $\iota_B$ invariant classes $k^i{\cal C}_i$ are characterised by
\begin{equation}
 k^1=k^5\;,\quad k^2=k^4\; ,
\end{equation}
so that $h^{1,1}_{\rm inv}(B)=4$. 

The associated CY three-fold has the following properties
\begin{align}
& h^{1,1}(X)=8 \;, \quad h^{2,1}(X)=80 \;, \quad \chi(X)=-144 \;, \\
& c_2(X)=4(C^0+C^{\hat{0}})+12(C^2+2C^3+C^4) \; ,
\end{align}
\bea
\mc{M}_X &=& \langle C^{\h{0}},C^0,-2C^1-C^2+C^4+2C^5+C^6,-C^0+C^1,C^2,C^4,-C^0+C^5,\nonumber\\ 
&& C^6, C^1+C^2+C^3-C^4-3C^5-2C^6,C^1-C^{\h{0}},C^5-C^{\h{0}} \rangle \, .
\eea

\subsection{Base $B=F_{15}$}
The basic topological characteristics of this space are
\begin{equation}
 K_B=-c_1(B)=-[x_3]-2[x_4]-2[x_5]-2[x_6]-[x_7]\;,\quad c_2(B)=8\;,\quad h^{1,1}(B)=6\; .
\end{equation}
A suitable choice for the basis of curve classes is given by
\begin{equation}
 {\cal C}^i=[x_{i+1}]\;,\quad i=2,\ldots ,7\; ,
\end{equation}
and the dual basis can be inferred from ${\cal C}_i=g_{ij}{\cal C}^j$ using the intersection numbers 
\be
(g_{ij})=
\left(\begin{array}{*{6}{@{}W{\mycolwd}@{}}}
-2 & -1 & 0 & 1 & 2 & 1 \\
-1 & -1 & 0 & 1 & 2 & 1 \\
0 & 0 & 0 & 1 & 2 & 1 \\
1 & 1 & 1 & 0 & 0 & 0 \\
2 & 2 & 2 & 0 & -2 & -1 \\
1 & 1 & 1 & 0 & -1 & -1 
\end{array}\right)\;,\quad
(\lambda_i)= ( 0, -1, -2, -2, -2, -1 )\;,\quad
\lambda=4\; .
\end{equation}
The Mori cone is generated by
\begin{equation}
 {\cal M}_B=\langle [x_2],\ldots ,[x_7],[x_2+x_3+x_4-x_6-x_7],[-2x_2-x_3+x_5+2x_6+x_7]\rangle\; .
\end{equation} 
For the involution, we find there are two inequivalent choices, which we refer to as cases (a) and (b). Their actions on the homogeneous coordinates are given by
\begin{eqnarray}
 \iota_B^{(a)}(x_1,\ldots ,x_8)&=&(x_5,x_6,x_7,x_8,x_1,x_2,x_3,x_4)\\
  \iota_B^{(b)}(x_1,\ldots ,x_8)&=&(x_7,x_6,x_5,x_4,x_3,x_2,x_1,-x_8)
\end{eqnarray}
with corresponding actions
\be
I_B^{(a)}=
\left(\begin{array}{*{6}{@{}W{\mycolwd}@{}}}
0 & 0 & 0 & 0 & 1 & 0 \\
0 & 0 & 0 & 0 & 0 & 1 \\
1 & 1 & 1 & 0 & -1 & -1 \\
-2 & -1 & 0 & 1 & 2 & 1 \\
1 & 0 & 0 & 0 & 0 & 0 \\
0 & 1 & 0 & 0 & 0 & 0
\end{array}\right)\;,\quad
I_B^{(b)}=
\left(\begin{array}{*{6}{@{}W{\mycolwd}@{}}}
0 & 0 & 1 & 0 & 0 & 0 \\
0 & 1 & 0 & 0 & 0 & 0 \\
1 & 0 & 0 & 0 & 0 & 0 \\
-2 & -1 & 0 & 1 & 2 & 1 \\
1 & 1 & 1 & 0 & -1 & -1 \\
0 & 0 & 0 & 0 & 0 & 1
\end{array}\right) \; ,
\ee
on the curve classes. The $\iota_B$ invariant curve classes $k^i{\cal C}_i$ are characterised by
\begin{equation}
 \begin{array}{lllllllll}
  k^1&=&k^5&\quad&k^2&=&k^6&\quad&\mbox{case (a)}\\
  k^1&=&k^3&\quad&k^2&=&k^6+2k^5-2k^3&\quad&\mbox{case (b)}
\end{array}\; ,
\end{equation}  
which means that $h^{1,1}_{\rm inv}(B)=4$ in either case.  

For the associated CY three-fold we find
\begin{align}
& h^{1,1}(X)=8 \;, \quad h^{2,1}(X)=80 \;, \quad \chi(X)=-144 \;, \\
& c_2(X)=4(C^0+C^{\hat{0}})+12(C^2+2C^3+2C^4+2C^5+C^6) \; ,
\end{align}
\bea
\mc{M}_X &=& \langle C^{\h{0}},C^0,-C^0+C^1+C^2+C^3-C^5-C^6,-2C^1-C^2+C^4+2C^5+C^6, \nonumber\\
&& -C^0+C^1,C^2,-C^0+C^3,C^4,-C^0+C^5,C^6,C^1+C^2+C^3-C^5-C^6-C^{\h{0}},\nonumber\\
&& C^1-C^{\h{0}},C^3-C^{\h{0}},C^5-C^{\h{0}} \rangle \;.
\eea

\section{Line bundles on elliptically fibered CY manifolds}
\label{linebundles_cohomologies_appendix}
In this section we collect useful information on line bundles on elliptically fibered CY manifolds, in particular on how to calculate their cohomology. The material of this appendix can also be found in the literature~\cite{Donagi:2004ia,Friedman:1997yq,Andreas:2007ev} and we collect and summarise all the properties relevant to heterotic line bundle model building. 

\subsection{Basic properties of line bundles}
We consider CY three-folds $X$ with a basis $\{D_I\}$ of divisor classes and a dual basis $\{C^I\}$ of curve classes (such that $D_I\cdot C^J=\delta_I^J$) and denote the triple intersection numbers by $d_{IJK}=D_I\cdot D_J\cdot D_K$. A general K\"ahler form is written as $J=t^ID_I$, where $t^I$ are the K\"ahler moduli relative to the chosen basis of divisor classes.

The line bundle $L$ on $X$ associated to the divisor $D=k^ID_I$ is denoted by $L={\cal O}_X(D)$ or sometimes, when it is clear which divisor basis is being referred to, simply by $L={\cal O}_X({\bf k})$. The first Chern class of this line bundle is given by
\be
 c_1(L)=D=k^ID_I\; .
\ee
For the Chern character one finds
\bea
{\rm ch}_1(L)&=&c_1(L)=k^ID_I \;, \\
{\rm ch}_2(L)&=&\frac{1}{2}c_1(L)^2=\frac{1}{2}d_{IJK}k^Ik^JC^K \;, \\
{\rm ch}_3(L)&=&\frac{1}{6}c_1(L)^3=\frac{1}{6}d_{IJK}k^Ik^Jk^K \; .
\eea
The Todd class of a CY three-fold is given by ${\rm Td}(X)=1+c_2(X)/12$, where $c_2(X)=c_{2I}(X)C^I$ is its second Chern class. Then, using the index theorem ${\rm ind}(L)={\rm ch}(L){\rm Td}(X)$, the index of a line bundle can be written as
\be
 {\rm ind}(L)={\rm ch}_3(L)+\frac{1}{12}c_1(L)c_2(X)=\frac{1}{6}d_{IJK}k^Ik^Jk^K+\frac{1}{12}k^Ic_{2I}(X)\; .
\ee
Another quantity we will require is the slope of the line bundle $L$, defined by
\be
 \mu_X(L):=\int_X J\wedge J\wedge c_1(L)=d_{IJK}t^It^Jk^K \, .
\ee

These above results for single line bundles can be easily generalised to line bundle sums
\begin{equation}
 V=\bigoplus_{a=1}^rL_a\quad\mbox{where}\quad L_a={\cal O}_X({\bf k}_a)\; ,
\end{equation}
with rank $r={\rm rk}(V)$. The first Chern class for such a line bundle sum is
\begin{equation}
  c_1(V)=\left(\sum_{a=1}^nk^I_a\right)D_I \; .
\end{equation}
We will normally be interested in line bundle sums $V$ with $c_1(V)=0$ and in this case the second Chern class and the index of $V$ are given by
\be
 c_2(V)=-\frac{1}{2}d_{IJK}\left(\sum_ak_a^Ik_a^J\right)C^K\, ,\quad
  {\rm ind}(V)=\frac{1}{6}d_{IJK}\sum_ak^I_ak^J_ak^K_a\, .
\ee

\subsection{Preliminaries for cohomology calculations}
\label{app:cohom_prelims}
In this appendix, we collect a number of mathematical results which will enter the calculation of line bundle cohomology.

We will make use of the \textbf{Serre duality theorem}. Let $X$ be a compact complex $n$-dimensional manifold with canonical bundle $K_X$, and let $V\rightarrow X$ be a vector bundle with dual bundle $V^*$. Then, Serre duality states that
\be
H^q(X,V) \cong H^{n-q}(X,K_X \otimes V^*)\,.
\ee
In particular, for a CY three-fold, the canonical bundle $K_X$ is trivial, so that $H^q(X,V) \cong H^{3-q}(X,V^*)$.

In order to exploit the fibration structure of the CY three-folds we will require \textbf{(higher) direct images}. Let $X$ be an elliptically fibered CY three-fold, with base $B$ and projection map $\pi : X \to B$. Given a short exact sequence of vector bundles, $V_i\rightarrow X$, one can ask if such a sequence implies the existence of a short exact sequence for the direct images (or push-forwards), $\pi_*V_i$, which are bundles on the base $B$. This question finds a general answer in the language of derived functors and higher direct images, as follows. Suppose we have the  short exact sequence of bundles
\be
0 \to V_1 \to V_2 \to V_3 \to 0 \; .
\ee
The direct image functor $\pi_*$ is left exact, so applying it to the above sequence leads to the exact sequence
\be
0 \to \pi_* V_1 \to \pi_* V_2 \to \pi_* V_3 \,.
\ee
How can this sequence be continued on the right, so as to keep it exact? The answer is (see, for example, Chapter III, Theorem 1.1A of Ref.~\cite{Hartshorne1977})
\bea
0 &\to& \pi_* V_1 \to \pi_* V_2 \to \pi_* V_3 \\
&\to& R^1\pi_* V_1 \to R^1\pi_* V_2 \to R^1\pi_* V_3 \\
&\to& R^2\pi_* V_1 \to R^2\pi_* V_2 \to R^2\pi_* V_3 \\
&\to& \ldots \,,
\eea
where $R^q\pi_*$ are called the `right derived functors' of the direct image functor, and $R^q\pi_*V_i$ are called the `higher direct images' of the bundles $V_i$. It can be shown (see, for example, Chapter III, Proposition 8.1 of \cite{Hartshorne1977}) that the higher direct image $R^q\pi_* V$ ($q\geq1$) is the sheaf associated to the pre-sheaf
\be
U \mapsto H^q\left(\pi^{-1}(U),V|_{f^{-1}(U)}\right) \,.
\ee

There are two very useful identities for higher direct images that will be helpful below. The first of these identities is the \textbf{projection formula},
\be
R^q\pi_*(\pi^*\mc{L}\otimes V)= {\cal L} \otimes R^q\pi_*V \,,
\ee
where ${\cal L}$ is a bundle on the base $B$ and $V$ a bundle on the total space (see, for example, Chapter III.8 in Ref.~\cite{Hartshorne1977}). 
The second identity, referred to as \textbf{relative duality}, is given by (see, for example, III.12 in Ref.~\cite{barth})
\be
R^1\pi_*V=(\pi_*V^*)^* \otimes K_B\,.
\ee
This result is frequently applied in the context of spectral cover bundles, as for example in Ref.~\cite{Friedman:1997yq}.

Finally, we will make extensive use of the \textbf{Leray spectral sequence} (see, for example, Chapter 5 of Ref.~\cite{weibel1994an}). This sequence relates cohomologies of bundles on the total space $X$ to cohomologies of bundles on the base $B$. For our purposes the following will suffice. For a bundle $V$, write $H^p \equiv  H^p(X,V)$ and define $E_2^{p,q} \equiv H^p(B,R^q\pi_* V)$. From the Leray spectral sequence one can show the existence of the  exact sequence
\be
0 \to E_2^{1,0} \to H^1 \to E_2^{0,1} \to E_2^{2,0} \to \mathrm{Ker}\left(H^2 \to E_2^{0,2}\right) \to E_2^{1,1} \to E_2^{3,0} \; .
\ee
In our case, $E_2^{3,0}$ is a third cohomology on the two-dimensional base $B$ and, hence, vanishes. Similarly, $E_2^{0,2}$ vanishes, since it involves $R^2\pi_* V$, and this is associated to second cohomologies on the one-dimensional fiber. As a result we have $\mathrm{Ker}\left(H^2 \to E_2^{0,2}\right) = H^2$ and the above exact sequence becomes
\be
0 \to E_2^{1,0} \to H^1 \to E_2^{0,1} \to E_2^{2,0} \to H^2 \to E_2^{1,1} \to 0 \,.
\ee
If this sequence splits (as it will in all cases we are interested in) it determines $H^1$ and $H^2$ in terms of cohomology on the base $B$. The other two cohomologies can also be expressed in terms of cohomology on the base, via the relations $H^0=E_2^{0,0}$ and $H^3=E_2^{2,1}$.

\subsection{Line bundle cohomology on the base}
\label{delpezzo_cohomology}

For most of the base spaces considered in this paper, no analytic expressions for line bundle cohomology are available, and we use the code \cohomCalg \cite{CohomOfLineBundles:Algorithm,cohomCalg:Implementation} to compute cohomologies on the base. However the cohomology of line bundles on del Pezzo surfaces has been computed, in Appendix B of Ref.~\cite{Blumenhagen:2008zz}. Since one of the base spaces we consider is a del Pezzo surface, we briefly present these results\footnote{Note however that the expression of their final result, equation (286), contains a typo: in $h^1$, the term $A_{\sum c_jp_j}(a)$ should appear with a plus rather than a minus.} for reference. Consider a del Pezzo surface $dP_r$ with $r\leq8$, given by a $\mbb{P}^2$ blown-up at $r$ distinct points $p_a$, where $a=1,\ldots ,r$, with hyperplane class $l$ and exceptional divisors $E_a=\rho^{-1}(p_a)$ and the blow-down map $\rho$. A line bundle on $dP_r$ can then be written as
\be
{\cal L} = \mc{O}_{dP_r}\left(nl+\sum_a b_aE_a-\sum_a c_aE_a\right)\,,
\ee
where $n\in\mbb{Z}$ and $b_a,c_a \in \mbb{Z}^{\geq 0}$.  For $n\geq-2$, the cohomology of the line bundle ${\cal L}$ is given by
\be
h^{q}(B,{\cal L})=
\begin{cases}
A_{\sum c_ap_a}(n) 															& \text{for }q=0 \\
A_{\sum c_ap_a}(n)-\binom{n+2}{2}+\sum_a \binom{b_a}{2}+\sum_a\binom{c_a+1}{2} 	& \text{for }q=1 \\
0 																		& \text{for }q=2 
\end{cases} \,,
\ee
while the cohomology for $n<-2$ can be obtained by Serre duality, $h^q(B,{\cal L})=h^{2-q}(B,{\cal L}^* \otimes K_{dP_r})$, using the expression
\begin{equation}
 K_B={\cal O}_{dP_r}\left(-3l+\sum_aE_a\right)
\end{equation}
for the canonical bundle.  Here $A_{\sum c_ap_a}(n)$ is the dimension of the space of homogeneous polynomials of degree $n$ on $\mbb{P}^2$ that have order $c_a$ zeroes at the points $p_a$.

\subsection{Line bundle cohomology for elliptic fibrations with a single section}

As a preparation, we first consider line bundle cohomology on an elliptically fibered CY three-fold $X$ over base $B$, with projection $\pi: X \to B$ and a single section $\sigma: B \to X$. To make some formulae easier to read, in this subsection and the next we will sometimes abuse notation and write $\sigma$ for the image of the section $\sigma(B)$, and similarly $\zeta$ for $\zeta(B)$, and it should be clear from context which one is meant. Let $V\rightarrow X$ be a vector bundle over $X$ and, as before, we write $H^p=H^p(X,V)$ and $E_2^{p,q}\equiv H^p(B,R^q\pi_*V)$. From Appendix \ref{app:cohom_prelims} we know that the Leray spectral sequence implies the exact sequence
\be
0 \to E_2^{1,0} \to H^1 \to E_2^{0,1} \to E_2^{2,0} \to H^2 \to E_2^{1,1} \to 0 \; ,
\ee
and, in addition, the results $H^0=E_2^{0,0}$ and $H^3=E_2^{2,1}$. A line bundle $L$ on $X$ can be written as
\be
L\equiv\mc{O}_X(n\sigma)\otimes\pi^*{\cal L}\; ,
\ee
where ${\cal L}$ is a line bundle on $B$ and $n$ is an integer. Using the projection formula, $R^q\pi_*(V \otimes \pi^*{\cal L})= R^q\pi_*V \otimes {\cal L}$, we can write the exact sequence explicitly as
\bea
0 &\to& H^1(B,\pi_*\mc{O}_X(n\sigma)\otimes {\cal L}) \to H^1(X,L) \to H^0(B,R^1\pi_*\mc{O}_X(n\sigma)\otimes {\cal L}) \nonumber\\
&\to& H^2(B,\pi_*\mc{O}_X(n\sigma)\otimes {\cal L}) \to H^2(Z,L) \to H^1(B,R^1\pi_*\mc{O}_X(n\sigma)\otimes {\cal L}) \to 0 \,. \label{lss}
\eea
To work this out further we require the (higher) direct images of the line bundles $\mc{O}_X(n\sigma)$. These have been worked out in Appendix C of Ref.~\cite{Donagi:2004ia} and the result is
\be
\pi_*\mc{O}_X(n\sigma)=
\begin{cases}
0														&\text{for}~ n<0  		\\
\mc{O}_B												&\text{for}~ n=0,1 		\\
\mc{O}_B\oplus K_B^2\oplus K_B^3\ldots\oplus K_B^{n} 		&\text{for}~ n\geq2  
\end{cases}\,,
\ee
\be
R^1\pi_*\mc{O}_X(n\sigma)=
\begin{cases}
0 																&\text{for}~ n>0  		\\
K_B 																&\text{for}~ n=-1,0  	\\
K_B^{1}\oplus K_B^{-1}\oplus K_B^{-2}\oplus\ldots\oplus K_B^{1-n} 		&\text{for}~ n\leq-2  		
\end{cases}\,.
\ee
From these results, we should distinguish the following cases.
\begin{itemize}
\item \underline{$n>0$}: We have $R^1\pi_*L=0$, and so,
\be
H^3=0\,, ~\text{and}~ H^p=E_2^{p,0} ~\text{for}~p=0,1,2\,.
\ee
\item \underline{$n<0$}: We have $\pi_*L=0$, and so,
\be
H^0=0\,, ~\text{and}~ H^p=E_2^{p-1,1} ~\text{for}~p=1,2,3\,.
\ee
\item \underline{$n=0$}: We have $\pi_*L={\cal L}$ and $R^1\pi_*L=K_B \otimes {\cal L}$, and so,
\be
H^0=E_2^{0,0}\,,~H^3=E_2^{2,1}\; .
\ee
In this case, the other two cohomologies, $H^1$ and $H^2$ can only be obtained by elementary methods if the sequence~\eqref{lss} splits, that is, if $E_2^{0,1}=0$ or $E_2^{2,0}=0$. Fortunately, this turns out to be always the case for the base spaces we consider.
\end{itemize}

\subsection{Line bundle cohomology for elliptic fibrations with two sections}\label{app:cohomtwosections}

As we have seen, elliptically fibered CY three-folds $X$ with two sections, $\sigma:B\rightarrow X$ and $\zeta:B\rightarrow X$, have to be blown-up and, as a result, contain a further curve and divisor class, relative to the single section case. The most general line bundle $L\rightarrow X$ can now be written as
\begin{equation}
 L={\cal O}_X(m\sigma)\otimes{\cal O}_X(n\zeta)\otimes\pi^*{\cal L}\; ,
\end{equation}
where $m,n\in\mathbb{Z}$ and  ${\cal L}\rightarrow B$ is a line bundle on the base. In fact, for our model building purposes we only require line bundles in a sub-class which consists of line bundles of the form
\begin{equation}
 L={\cal O}_X(n\Sigma)\otimes \pi^*{\cal L}\; , \label{Lform}
\end{equation}
where $n\in\mathbb{Z}$ and $\Sigma=\sigma(B)+\zeta(B)$. 

In order to compute the cohomology of such line bundles, we follow Appendix A of Ref.~\cite{Andreas:2007ev}. Note that the subsequent derivation only holds if $H^1(B,K_B^{-n})=0$ for all $n\geq0$, a condition which is satisfied for toric weak Fano base spaces $B$ and, hence, for all base spaces considered in this paper. That this condition holds for toric weak Fano bases can be seen as follows (for more information see for example Ref.~\cite{cox2011toric}). The toric Kawamata-Viehweg theorem states that on a compact toric variety $B$, if $D$ is a nef and big divisor then
\be
H^p(B,K_B \otimes \mathcal{O}_B(D)) = 0 \quad \textrm{for all $p>0$} \; .
\ee
If $B$ is weak Fano, then $K_B^{-1}$ is of the form $\mathcal{O}_B(D)$ where $D$ is nef and big, and from this it follows that this is true also of $K_B^{-n}$ for $n>1$. Hence for weak Fano bases $B$, using $K_B^{-n-1}$ in the theorem we have that $H^1(B,K_B^{-n}) = 0$ for $n\geq0$.

We begin by recalling relative duality which states that
\be
R^1\pi_*L=(\pi_*L^*)^* \otimes K_B \; .
\ee
In particular, this implies that $R^1\pi_*\mc{O}_X(n\Sigma)=0$ for $n>0$. Now consider the short exact sequence,
\be
0 \to \mc{O}_X(-\zeta) \to \mc{O}_X \to \mc{O}_{\zeta} \to 0 \,,
\ee
and tensor this sequence with $\mc{O}_X(\Sigma)$ to obtain
\be
0 \to \mc{O}_X(\sigma) \to \mc{O}_X (\Sigma)\to \mc{O}_X(\Sigma) \otimes \mc{O}_{\zeta} \to 0 \,.
\ee
Now we consider the associated long exact sequence of higher direct images, but, since $R^1\pi_*\mc{O}_X(\sigma)=0$, this leads to the short exact sequence
\be
0 \to \mc{O}_B \to \pi_*\mc{O}_X (\Sigma)\to K_B \to 0 \,, \label{splitseq}
\ee
for the direct images. Here, we have used that $\pi_*\mc{O}_X(\sigma)=\mc{O}_B$ and $\pi_*(\mc{O}_X(\Sigma) \otimes \mc{O}_{\zeta})=K_B$. The space of extensions defined by the sequence~\eqref{splitseq} is isomorphic to $H^1(B,K_B^*)$ and, from our assumption about the base space $B$, this cohomology vanishes. As a result, the sequence~\eqref{splitseq} splits and we have
\be
\pi_* \mc{O}_X(\Sigma)=\mc{O}_B \oplus K_B\;.
\ee
Starting from this result, we will now prove by induction that
\be
\pi_*\mc{O}_X(n\Sigma)=\mc{O}_B \oplus K_B \oplus \left(K_B^{\otimes2} \oplus K_B^{\otimes3} \oplus \ldots \oplus K_B^{\otimes n}\right)^{\oplus2} \quad \text{for all }n>1\; . \label{pfres}
\ee
The starting point for the proof is the short exact sequence
\be
0 \to \mc{O}_X((n-1)\Sigma) \to \mc{O}_X(n\Sigma) \to \mc{O}_X(n\Sigma) \otimes \mc{O}_X|_{\Sigma} \to 0\; ,
\ee
along with the relation $\mc{O}_X(n\Sigma) \otimes \mc{O}_X|_{\Sigma}=\mc{O}_X(n\Sigma)|_{\Sigma}$. Since $R^1\pi_*\mc{O}_X(m\Sigma)=0$ for all $m>0$, we have $R^1\pi_*\mc{O}_X((n-1)\Sigma)=0$ for $n>1$, so that the associated long exact sequence of higher direct images truncates to the short exact sequence
\be
0 \to \pi_*\mc{O}_X((n-1)\Sigma) \to \pi_*\mc{O}_X(n\Sigma) \to \pi_*\mc{O}_X(n\Sigma)|_{\Sigma} \to 0\;.
\ee
We note that $\pi_*\mc{O}_X(n\Sigma)|_{\Sigma}=K_B^n \oplus K_B^n$. Recall that we are working with a base $B$ for which $H^1(B,K_B^{-m})=0$ for all $m\geq0$. Then, by the induction hypothesis for $\pi_*\mc{O}_X((n-1)\Sigma)$, we see that this exact sequence splits, and that the result~\eqref{pfres} follows. An expression for $R^1\pi_*\mc{O}_X(n\sigma)$ can then be obtained from relative duality.

In summary, the results for the direct images and higher direct images of the line bundles $\mc{O}_X(n\Sigma)$ for all $n\in\mbb{Z}$, where $\Sigma \equiv \sigma(B) + \zeta(B)$, are:
\be
\pi_*\mc{O}_X(n\Sigma)=
\begin{cases}
0																	&\text{for}~ n<0  		\\
\mc{O}_B															&\text{for}~ n=0 		\\
\mc{O}_B\oplus K_B													&\text{for}~ n=1 		\\
\mc{O}_B \oplus K_B \oplus \left(K_B^{\otimes2} \oplus K_B^{\otimes3} \oplus \ldots \oplus K_B^{\otimes n}\right)^{\oplus2} 									&\text{for}~ n\geq2  
\end{cases}\,,
\ee
\be
R^1\pi_*\mc{O}_X(n\Sigma)=
\begin{cases}
0 																	&\text{for}~ n>0  		\\
K_B 																	&\text{for}~ n=0  		\\
\mc{O}_B\oplus K_B													&\text{for}~ n=-1 		\\
\mc{O}_B \oplus K_B \oplus \left(K_B^{\otimes(-1)} \oplus K_B^{\otimes(-2)} \oplus \ldots \oplus K_B^{\otimes(-n+1)}\right)^{\oplus2}  	 							&\text{for}~ n\leq-2  		
\end{cases}\,.
\ee
Deriving the cohomology for line bundles $L$ of the form~\eqref{Lform} then proceeds in complete analogy with the single section case discussed in the previous sub-section. All we have to do is replace $\sigma$ by $\Sigma$ and, instead of using the results for the (higher) direct images of ${\cal O}_X(n\sigma)$, we use the above results for ${\cal O}_X(n\Sigma)$.

\bibliographystyle{JHEP}
\providecommand{\href}[2]{#2}\begingroup\raggedright

\endgroup

\end{document}